\documentclass[twocolumn]{aastex62}

\usepackage{epsfig}
\usepackage{epstopdf}
\usepackage{graphics}
\usepackage{amsmath}
\usepackage{xcolor}
\usepackage{aas_macros}

\usepackage{array}
\usepackage{booktabs}
\usepackage{amsmath,amssymb,amsfonts,amsbsy}
\usepackage{mathrsfs}
\usepackage{url}
\usepackage{multirow}
\usepackage{xcolor}
\usepackage{color}
\usepackage{ulem}
\usepackage{enumerate}

\renewcommand{\vec}[1]{\boldsymbol{#1}}

\newcolumntype{Y}{D{,}{\pm}{-1}}

\renewcommand{\figurename}{Fig.}
\renewcommand{\tablename}{Tab.}

\renewcommand{\arraystretch}{1.5}

\begin{document}
\setlength{\tabcolsep}{20pt}


\title{The Compton-Getting origin of the large-scale anisotropy of Galactic cosmic rays}

\correspondingauthor{Wei Liu, Huirong Yan, Yi-qing Guo }
\email{liuwei@ihep.ac.cn, huirong.yan@desy.de, guoyq@ihep.ac.cn}

\author{Bing-Qiang Qiao}
 \affiliation{Deutsches Elektronen Synchrotron DESY, Platanenallee 6, D-15738, Zeuthen, Germany}
 \affiliation{Institut für Physik und Astronomie, Universität Potsdam, D-14476, Potsdam, Germany}
\author{Wei Liu}
\affiliation{Key Laboratory of Particle Astrophysics, Institute of High Energy Physics, Chinese Academy of Sciences, Beijing 100049, China}
\author{Huirong Yan}
 \affiliation{Deutsches Elektronen Synchrotron DESY, Platanenallee 6, D-15738, Zeuthen, Germany}
 \affiliation{Institut für Physik und Astronomie, Universität Potsdam, D-14476, Potsdam, Germany}
\author{Yi-qing Guo}
\affiliation{Key Laboratory of Particle Astrophysics, Institute of High Energy Physics, Chinese Academy of Sciences, Beijing 100049, China}
\affiliation{University of Chinese Academy of Sciences, Beijing 100049, China}
\affiliation{TIANFU Cosmic Ray Research Center, Chengdu 610000, China}


\begin{abstract}
Recent studies suggest that the anisotropy in cosmic-ray arrival directions can provide insight into local acceleration sites and propagation conditions. We developed a unified framework to interpret both the observed energy spectra and the large-scale anisotropy. In this work, we explore the influence of the Sun’s motion relative
to the local plasma frame—the Compton-Getting (CG) effect—on the anisotropy. We find that incorporating the CG effect could slightly reduce the dipole amplitude and shift the phase away from the direction of the local regular magnetic field at tens of TeV. At lower energies, where the anisotropy from the cosmic-ray density 
gradient is weak, the Sun’s relative motion becomes more prominent. Below $\sim 200$ GeV, the dipole amplitude 
increases again, approaching the value expected from the CG effect. Additionally, a phase flip is observed at a few hundred GeV, aligning with the CG direction. Future anisotropy measurements from $100$ GeV to TeV energies could serve as a critical test of this effect. 
\end{abstract}
 

\section{Introduction}
\label{sec:intro}
The origin and propagation of Galactic cosmic rays (CRs) below PeV energies remain unresolved questions in astrophysics. Observational data on CRs include their energy spectra, composition, and arrival directions. 
While earlier studies primarily focused on energy spectra and composition \citep{2007ARNPS..57..285S}, cosmic-ray (CR) anisotropy has attracted growing attention in recent years. Due to frequent scattering by interstellar magnetohydrodynamic (MHD) 
turbulence, the arrival directions of CRs are nearly isotropic \citep{1966ApJ...146..480J, 2002cra..book.....S, 
Yan2022icrc}. However, numerous measurements have revealed a nonuniform distribution of arrival directions, known as anisotropy \citep{2006Sci...314..439A, 2007PhRvD..75f2003G, 2008PhRvL.101v1101A, 2009ApJ...698.2121A, 
2010ApJ...718L.194A, 2010ApJ...711..119A, 2011ApJ...740...16A, 2012ApJ...746...33A, 2013ApJ...765...55A, 
2013PhRvD..88h2001B, 2014ApJ...796..108A, 2015ApJ...809...90B, 2016ApJ...826..220A, 2017ApJ...836..153A, 
2017ApJ...842...54L, 2019ApJ...871...96A}. The intensity of anisotropy varies between $\sim 10^{-4}$ and $\sim 10^{-3}$ in the TeV–PeV range and exhibits energy-dependent angular distributions.

Despite the fact that the CR trajectories are highly tangled by the interstellar magnetic field, abundant studies have shown that anisotropy can still retain some information of CR sources. Within the framework of the traditional diffusion model, the uneven distribution of overall CR sources predicts a large-scale anisotropy, with the direction aligning with the Galactic center immutably\citep{2012JCAP...01..011B, 
2017PhRvD..96b3006L}. If a local source contributes significantly to the CR flux, anisotropy may indicate its presence \citep{2017JCAP...01..006M, 2022ApJ...930...82L}. In recent works, we developed a unified scenario that incorporates a local source within a spatially dependent propagation model to explain both the energy spectrum and the dipole anisotropy \citep{2019JCAP...10..010L, 2019JCAP...12..007Q, 2023ApJ...942...13Q}. This model suggests that the excess of nuclei between 200 GeV and $\sim 20$ TeV, along with the energy-dependent anisotropy below 100 TeV, implies the presence of nearby CR sources. The supernova
remnant (SNR) at the birthplace of the Geminga pulsar could be a plausible candidate for such a local cosmic ray accelerator\citep{2022ApJ...926...41Z}, with similar conclusions supported by other recent studies \citep{2022MNRAS.511.6218Z, 2023ApJ...952..100N}.

The anisotropy also provides indirect insight into the local interstellar environment. Observations of energetic neutral atom emissions by the IBEX mission reveal a local regular magnetic field (LRMF) within 0.1 pc of the solar system, oriented at $l \simeq 210.5^\circ$, $b \simeq -57.1^\circ$ \citep{2009Sci...326..959M}. This
direction closely matches the observed dipole phase below 100 TeV. Some studies suggest that anisotropic diffusion guided by the LRMF directs CRs into the heliosphere \citep{2014Sci...343..988S, 2016PhRvL.117o1103A, 
2020ApJ...892....6L}. Furthermore, the energy-dependent of the dipole phase implies a transition from anisotropic to isotropic diffusion around 100 TeV in the local interstellar environment\citep{2024ApJ...962...43L}. Moreover, quadrupole anisotropies may be induced by nonuniform convection effects \citep{2022ApJ...938..106Z}, while smaller-scale anisotropies are likely attributed to the local turbulence on the scale of a few mean free paths, or heliospheric influences. For a comprehensive review, one can refer to \cite{2017PrPNP..94..184A}.

Notably, the non-uniform distribution of the arrival directions of CRs was predicted as early as the 1930s by \cite{1935PhRv...47..817C}. This is caused by the relative motion of the solar system through the local plasma frame. As the Sun has an average circular speed $v \simeq 220$ km/s around the Galactic center, it induces a dipole anisotropy, which is similar to the Doppler effect. The amplitude of this anisotropy is related to the power index of the CR spectrum and is expected to be $2.3 \times 10^{-3}$ with its peak at $\alpha \simeq 315^\circ$, $\delta \simeq 0^\circ$. This is well known as the Compton-Getting (CG) effect, denoted as $\rm CG_{GAL}$.
However, the expected dipole anisotropy is inconsistent with the measurement at $\sim 300$ TeV \citep{2006Sci...314..439A}. Thereupon, it is inferred that the Galactic CRs co-rotate with the local Galactic magnetic field environment. Nevertheless, another CG effect ($\rm CG_{SUN}$), which occurs in local solar time, has been detected.
It originates from the evolution of the Earth around the Sun \citep{1986Natur.322..434C, 2004PhRvL..93f1101A}. Since the revolution speed of the Earth is $\sim 30 ~$km/s, the amplitude is $~ 4.5 \times 10^{-4}$ and points to the local $6$ hour a.m. For the ultra-high energy cosmic rays, most of them are regarded of extragalactic origin. It has been argued that the motion of the Sun relative to the cosmic microwave background (CMB) frame could induce a dipole anisotropy with intensity $\sim 0.6 \%$, similar to that seen in the CMB \citep{2006PhLB..640..225K}.

In this work, we further investigate the impact of the CG effect, including $\rm CG_{LSR}$ and $\rm CG_{ISM}$, on CR anisotropy. Although CRs seem to corotate with the average motion of the stars in the whole Galaxy, the motion of the Sun relative to the local plasma frame is considered. Since the plasma rest frame is not precisely defined, we explore two scenarios: one is the so-called local standard of rest (LSR), which corresponds to the Sun's motion towards the solar apex, and the other is the relative motion through the local interstellar medium(ISM). We find that in both cases, below 10 TeV, the dipole phase no longer points towards the direction of the local regular magnetic field at the lower energy as the anisotropy from the CR gradient gradually decreases. Our predictions can be tested by future anisotropy measurements at sub-TeV energies\citep{2019arXiv190502773C,2024NuScT..35..149P}.

\section{Model Description}
\label{sec:model}

\subsection{Spatially-dependent propagation}
\label{subsec:SDP}

In this study, we employ the spatially dependent propagation (SDP) model to describe Galactic cosmic-ray transport. This class of models has been developed and applied extensively over the past decades, providing a unified framework for interpreting diverse observational results, such as the hardening of primary and secondary spectra, diffuse $\gamma$-ray emission, and anisotropy at large scales.
\citep{2012ApJ...752L..13T, 2015PhRvD..92h1301T, 2016PhRvD..94l3007F, 2016ApJ...819...54G, 2018ApJ...869..176L, 
2018PhRvD..97f3008G, 2019JCAP...10..010L, 2019JCAP...12..007Q, 2020ChPhC..44h5102T, 2020FrPhy..1624501Y, 
2022ApJ...926...41Z}. Recent observations of TeV halos surrounding pulsars \citep{2017Sci...358..911A, 
2021PhRvL.126x1103A, 2023A&A...673A.148H, 2024ApJ...974..246A}, along with evidence of anisotropic diffusion 
in strongly magnetized turbulence (i.e., with $M_A \simeq \delta B/B_0 \leq 1$) \citep{2008ApJ...673..942Y, 2017JCAP...10..019C, 2020ApJ...892....6L, 2022ApJ...926...94M}, also strongly support the necessity of incorporating spatial dependence into the diffusion process, as demonstrated earlier from the perspective of inhomogeneous turbulence driving and damping \citep{2002PhRvL..89B1102Y, 2004ApJ...614..757Y, 2014ApJ...782...36E, 2020NatAs...4.1001Z, 2020PhRvX..10c1021M, Hou2025,Zhao2025}.
 
In the SDP picture, the Galactic halo is conceptually divided into two transport layers. The inner halo (IH), extending around the disk, is strongly affected by supernova-driven turbulence, which suppresses diffusion in regions of high source density. Conversely, in regions with relatively few sources, turbulence levels are lower and diffusion proceeds more efficiently. As a result, the diffusion coefficient in the IH depends on the radial distribution of CR sources. In contrast, the outer halo (OH), located farther from the Galactic disk, is less affected by stellar activity, and the diffusion coefficient there depends only on particle rigidity.

We define the total half-thickness of the diffusive halo as $z_h$, with the IH and OH having thicknesses 
$\xi z_h$ and $(1 - \xi) z_h$, respectively. Both $z_h$ and $\xi$ are determined by fitting the B/C ratio and the spectra of CR nuclei. The diffusion coefficient $D_{xx}$ is parameterized as: 
\begin{equation}
D_{xx}(r,z, {\cal R} )= D_{0}F(r,z)\beta_0^{\eta} \left(\dfrac{\cal R}{{\cal R}_{0}} \right)^{\delta_0 F(r,z)} ~,
\label{eq:diffusion}
\end{equation}
where ${\cal R}_{0}$ is the reference rigidity fixed at 4 GV, $\beta_0$ is the particle velocity normalized to the speed of light, $D_0$ and $\delta_0$ are constants representing the diffusion coefficient and its rigidity-dependent in the OH, and $\eta$ is a phenomenological parameter introduced to fit low-energy data. The spatial function $F(r, z)$ is given by: 
\begin{equation}
F(r,z) =
\begin{cases}
g(r,z) +\left[1-g(r,z) \right] \left(\dfrac{z}{\xi z_h} \right)^{n} , &  |z| \leqslant \xi z_h \\
1 ~, & |z| > \xi z_h
\end{cases},
\end{equation}
where $g(r,z) = N_m/[1+f(r,z)]$, with $f(r,z)$ representing the spatial distribution of CR sources. The CR sources are assumed to be 
axisymmetrically distributed as $f(r,z) \propto (r/r_\odot)^{\alpha} \exp[-\beta (r -r_\odot)/r_\odot] \exp(-|z|/z_s)$, where $r_\odot = 8.5$ kpc and $z_s = 0.2$ kpc. $\alpha$ and $\beta$ are taken as $1.69$ and $3.33$ \citep{1996A&AS..120C.437C}.

The injection spectrum of background sources is assumed to have a form of power-law of rigidity with an exponential cutoff, namely
\begin{equation}
Q({\cal R}) \propto {\cal R}^{-\nu} \exp \left(-\dfrac{{\cal R}}{{\cal R}_{\rm c} } \right) ~.
\end{equation}
The diffusion-reacceleration (DR) model is adopted in the propagation equation and the numerical package, DRAGON, is used to compute the background CR distribution \citep{1475-7516-2008-10-018}.

\subsection{Local source}

Building on earlier analyses \citep{2019JCAP...10..010L, 2019JCAP...12..007Q, 2022ApJ...926...41Z}, we consider the Geminga supernova remnant (SNR) as the likely origin of the spectral hardening above a few hundred GV and of the dipole anisotropy observed below $\sim$100 TeV. The age of the remnant, estimated from the spin-down evolution of the Geminga pulsar, is about $\tau = 3.4 \times 10^5$ years \citep{2005AJ....129.1993M}. Its location is determined to be $l = 194.3^\circ, b = -13^\circ$, at a distance of approximately 330 pc from the Sun \citep{2007Ap&SS.308..225F}. The contribution of this nearby source is modeled by solving the time-dependent diffusion equation using the Green’s function formalism, with free-escape conditions at large distances \citep{2017PhRvD..96b3006L, 2019JCAP...10..010L}. For the instantaneous and point-like injection, the CR density of local source can be calculated as
\begin{equation}
\psi(r,{\cal R},t)=\frac{q_{\rm inj}({\cal R})}{(\sqrt{2\pi}\sigma)^3}
\exp\left(-\frac{(\vec{r}-\vec{r}^\prime)^2}{2\sigma^2}\right),
\end{equation}
where $\sigma({\cal R},t)=\sqrt{2D_{xx}({\cal R})t}$ is the effective diffusion length within time $t$. The injection spectrum $q_{\rm inj}({\cal R})$ is parameterized as a power-law function of rigidity with an exponential cutoff, i.e. $q_0{\cal R}^{-\alpha} \exp(-{\cal R}/{\cal R}'_{\rm c})$. Note that the Geminga nearby source, which located at a specific position in the inner halo, is very close to the Solar System. Hence, its diffusion coefficient is taken as the local value at the Solar System given in Eq.~(\ref{eq:diffusion}) here.

Based on the injection spectrum parameters listed in Tab.~\ref{tab:para_inj} and the formula $W_{p,He}= \int q_{inj}(E) E dE$, we obtain total injection energies of about $1.5 \times 10^{50}$ erg for protons and $6.8 \times 10^{49}$ erg for helium from nearby source, corresponding to roughly 5-15\% of a canonical supernova explosion energy of $10^{51}$ erg, consistent with standard cosmic-ray acceleration efficiencies. For the background component, the estimated energy injection that includes all Galactic cosmic-ray sources except Geminga is approximately $6.5 \times 10^{50}$ erg, which is physically comparable to the injection energy from the nearby source. 
Since the background and nearby source components are computed using the numerical software(DRAGON) and the analytical Green’s function method, respectively, different units are therefore required to ensure dimensional consistency here. 

Why do we not include other nearby sources such as Monogem, Cygnus Loop, and Vela?
Because their contributions to the observed proton spectrum are strongly suppressed by their age-distance combinations. To illustrate this point, we separately computed the proton spectra from these nearby sources. Their corresponding parameters are summarized in Tab.~\ref{tab:srcparas}. For Monogem, the injection energy is scaled according to its X-ray–inferred explosion energy of $\sim$ \(1.9\times10^{50}\)~erg \citep{1996ApJ...463..224P}, which is about one order of magnitude lower than that of Geminga. For Vela and Cygnus Loop, the injection energies are assumed to be comparable to that of Geminga to facilitate a consistent comparison. These parameter selections are guided by observational and physical constraints, ensuring that the model predictions are well grounded rather than arbitrary. Under the same transport setup, our calculations show that the predicted proton fluxes from these nearby sources at Earth remain sub-dominant compared to that of Geminga. For relatively young sources, low-energy particles have not yet had sufficient time to diffuse to the Solar system, while for the more distant ones, the local flux is significantly suppressed.
A systematic study by \citet{2022ApJ...930...82L} also demonstrated that, under these considerations, Geminga stands out as the most plausible nearby contributor, while the other candidates are either negligible or inconsistent with current spectral and anisotropy data.

\begin{table*}[htbp]
\centering
\caption{Characteristic and injection parameters for other three nearby SNRs.}
\begin{tabular}{lccccc}
\hline
\hline
SNR Name & Distance (pc) & Age (yr) & ~~~$q_0^p~ [\rm GeV^{-1}]$ & ~~~$\alpha$~~~ & ~~~${\cal R}'_c$ [TV]~~~ \\
\hline
Monogem     & 288 & $8.6 \times 10^{4}$ & $3.2 \times 10^{51}$ & 2.10 & 30 \\
Vela     & 300 & $1.1 \times 10^{4}$ & $3.2 \times 10^{52}$ & 2.10 & 30 \\
Cygnus Loop & 720 & $2.0 \times 10^{4}$ & $3.2 \times 10^{52}$ & 2.10 &  30 \\
\hline
\end{tabular}
\label{tab:srcparas}
\end{table*}

\subsection{Anisotropic diffusion}

Observations with IBEX reveal a local regular magnetic field (LRMF) in the solar neighborhood with a characteristic spatial scale of order $\sim 0.1$ pc and orientation $l \simeq 210.5^\circ, b \simeq -57.1^\circ$ \citep{2009Sci...326..959M, 2013ApJ...776...30F}. Starlight polarization measurements independently support a consistent field direction \citep{2015ApJ...814..112F}, which is broadly aligned with the measured dipole phase below $\sim$100 TeV \citep{2016PhRvL.117o1103A}. Because the spatial scale of the LRMF ($\sim$0.1 pc) is much shorter than the typical propagation scale of Galactic CRs inferred from the B/C ratio, its influence on the energy spectra is negligible. In other words, the large-scale spectral features are governed primarily by transport over kiloparsec distances and are insensitive to magnetic-field structures on sub-parsec scales. Nevertheless, such a field can have a pronounced effect on the angular distribution of arriving particles, since the local anisotropy is shaped by the most recent segment of the propagation history, where the guiding effect of the LRMF can redirect particle trajectories without altering their overall energy distribution. In this scenario, the scalar diffusion coefficient $D_{xx}$ is replaced by a diffusion tensor $D_{ij}$, expressed as: 
\begin{equation} D_{ij} = D_\perp \delta_{ij} + (D_\parallel - 
D_\perp) b_i b_j ~, \quad b_i = \dfrac{B_i}{|\vec{B}|} \label{eq:D_ij_1} ~,
\end{equation}
where $b_i$ is the $i$-th component of the LRMF unit vector \citep{1999ApJ...520..204G, 2017JCAP...10..019C}. The diffusion coefficients parallel and perpendicular to the LRMF are given by: 
\begin{align}
D_\parallel &= D_{0\parallel} \beta_0^{\eta} \left(\frac{\cal R}{{\cal R}_0} \right)^{\delta_\|} ~, \\
D_\perp &=\,D_{0\perp} \beta_0^{\eta} \left(\frac{\cal R}{{\cal R}_0} \right)^{\delta_\perp} = \varepsilon D_{0\parallel} \beta_0^{\eta} \left(\frac{\cal R}{{\cal R}_0} \right)^{\delta_\perp} ~,
\label{eq:DparaDperp}
\end{align}
where $D_\parallel = D_{xx}$, i.e., $D_{0\parallel} = D_{0}F(r,z)$ and $\delta_\parallel = \delta_0 F(r,z)$. $\varepsilon = D_{0\perp}/D_{0\parallel}$ is the ratio between perpendicular and parallel diffusion coefficient at the reference rigidity ${\cal R}_0$. The energy-dependent of the ratio between perpendicular and parallel diffusion coefficients may be attributed to the difference in the path volumes that particles of different energy go through. The higher energy particles propagating through the outer or inner halos before reaching the solar system are like to traverse a larger path volume, which includes more sources, and thus experience stronger perturbations. In contrast, low-energy particles travel through a smaller path volume and naturally undergo weaker disturbances. This could lead to an energy-dependent in this ratio. Taking local anisotropic diffusion into account, the dipole anisotropy becomes: 
\begin{equation} 
\vec{\Delta_{grad}} 
= \dfrac{3\vec{D} \cdot \nabla \psi}{v \psi} = \dfrac{3}{v \psi} D_{ij} \dfrac{\partial \psi}{\partial x_j}. 
\label{eq:ampgrad}
\end{equation}

\subsection{Compton-Getting effect}
Furthermore, the motion of the Solar System through the Galaxy introduces an energy-independent dipole anisotropy due to the CG effect \citep{1935PhRv...47..817C}: 
\begin{align}
\vec{\Delta}_{\rm CG} = (2+\gamma)\,\dfrac{\vec{v}_\odot}{c} ~,
\label{eq:ampcg}
\end{align}
where $v_\odot$ is the velocity of the Sun with respect to the LSR or ISM frame and correspondingly $\Delta_{\rm CG}$ stands for either $\rm CG_{LSR}$ or $\rm CG_{ISM}$. $\gamma$ is the CR spectral index \citep{1970P&SS...18...25F}. In practice, the total dipole anisotropy is obtained as the vector sum of the gradient-driven contribution [Eq.~(\ref{eq:ampgrad})] and the CG term [Eq.~(\ref{eq:ampcg})]:
\begin{equation}
\vec{\Delta}_{\rm tot} = \vec{\Delta}_{\rm grad} + \vec{\Delta}_{\rm CG}.
\end{equation}
The velocity is initially thought to be the circular speed of the Sun around the Galactic center, which is approximately $220$ km/s. The dipole anisotropy is expected to be $2.3 \times 10^{-3}$ with its peak at $\alpha \simeq 315^\circ, \delta \simeq 0^\circ$. However, this expectation is inconsistent with YBJ experiment results at $\sim 300$ TeV, suggesting that CRs co-rotate with the average stellar motion.

Moreover, the exact rest frame of the cosmic rays is ambiguous. Many observations have shown that the Sun is not stationary relative to the average motion of the local stars and the local interstellar medium, and its relative speed is smaller than $220$ km/s. If the overall cosmic rays remain at rest with the average motion of the local stars or the local interstellar medium, the relative motion of the Sun could induce a dipole anisotropy. Two kinds of candidates can be considered for the local rest reference frame. One is the so-called local standard of rest (LSR). 
It is defined as the reference frame moving with the average velocity of stars in the solar neighborhood, including the Sun, and has no intrinsic spatial extent \citep{2010MNRAS.403.1829S}. 
In practice, observational determinations of the LSR are based on stellar samples in the solar neighbourhood, typically spanning distances from roughly 100~pc to several hundred parsecs from the Sun. \citep{2021A&A...649A...6G, 2024A&A...687A.286Z}. 
This empirical scale thus remains applicable to our analysis. By the local standard of rest, the space velocities of stars average out to zero. The LSR is moving around the center of the Galaxy at a speed of approximately $200 \sim 240$ km/s. The motion of the Sun relative to the LSR is found to be $v_{\rm LSR} \approx 18.0 \pm 0.9 ~ \rm km/s$, which points to $l \approx 47.9^\circ \pm 2.9^\circ$ and $b \approx 23.8^\circ \pm 2.0^\circ$ \citep{2010MNRAS.403.1829S}. Another possibility would be a relative velocity through the local interstellar medium (ISM). The velocity vector measured by the IBEX Experiment is $v_{\rm ISM} \approx 23.2 \pm 0.3 ~\rm km/s$ with orientation $l \approx 5.25^\circ \pm 0.24^\circ$ and $b \approx 12.0^\circ \pm 0.5^\circ$ \citep{2012Sci...336.1291M}. It's worth noting that the IBEX-inferred flow direction here is obtained indirectly from charge-exchange measurements of interstellar neutral atoms. Therefore, by simultaneously considering the LSR and the IBEX-based LISM reference frames, we aim to make the evaluation of the CG effect as comprehensive and reliable as possible within the current uncertainties. Furthermore, current observations show no compelling evidence for a substantial bulk motion of the Geminga SNR. Therefore, its motion relative to the local stellar population or the surrounding ISM is neglected in our calculations.

\section{Results and Discussion}
\label{sec:results}
 
\begin{figure*}[!ht]
	\centering
	\includegraphics[width=0.55\textwidth]{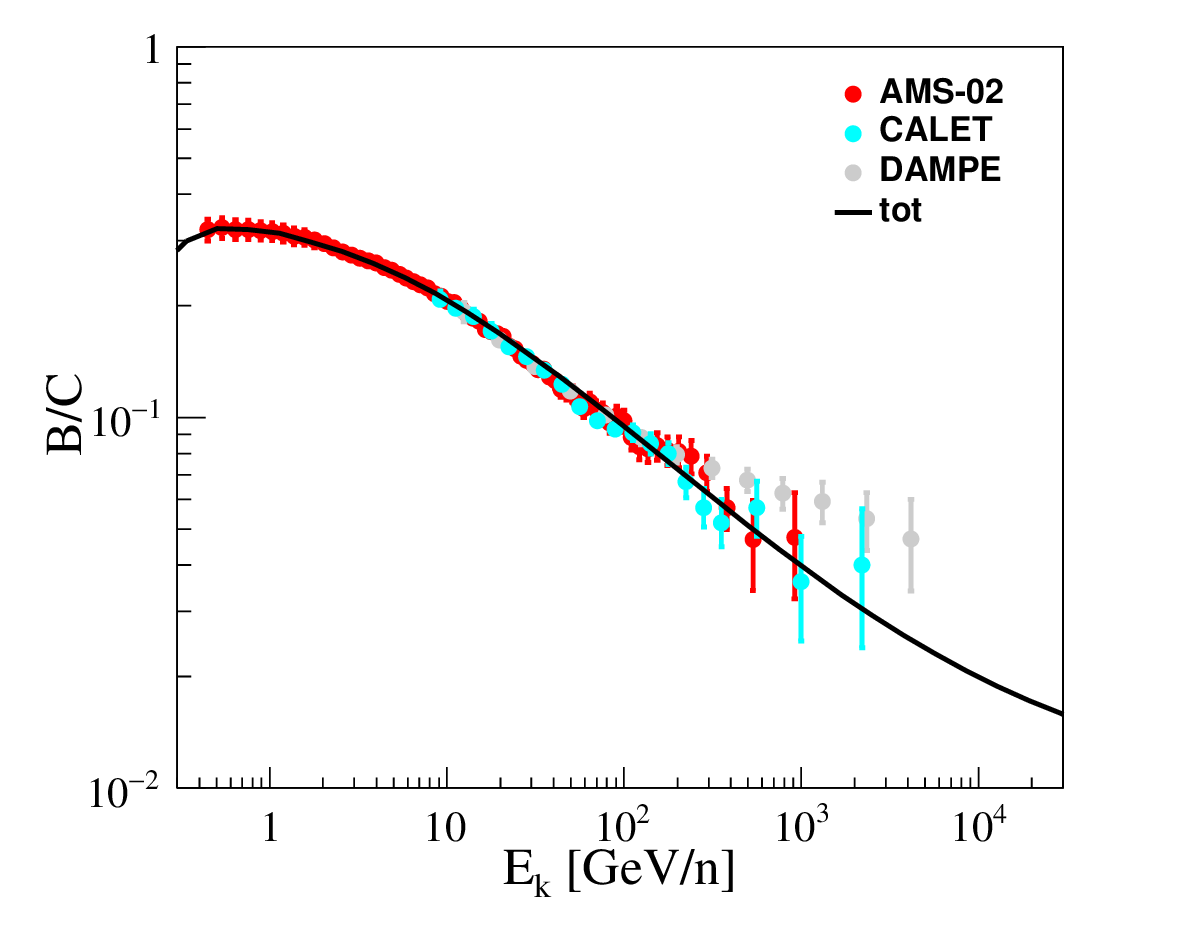}
	\caption{Comparison between model calculations and observations for B/C ratio. The data points are adopted from AMS-02, CALET and DAMPE measurements \citep{2016PhRvL.117w1102A, 2022PhRvL.129y1103A, 2022SciBu..67.2162D}.
	}
	\label{fig:BCratio}
\end{figure*}

\begin{deluxetable}{lll}[tb!]
	\tablecolumns{2}
	\tablewidth{0.5\textwidth}
	\tablecaption{Fitted propagation parameters of SDP.
		\label{tab:para_trans}}
	\tablehead{
		\colhead{Parameter} & & \colhead{Value} 
	}
	\startdata
	$D_0$ [$\mathrm{cm}^2\mathrm{s}^{-1}$] &  & $4.80 \times 10^{28}$  \\
	$\delta_0$ &  & 0.56 \\
	$N_{\rm m}$ &  & 0.58 \\
	$\xi$ &  & 0.1 \\
	$n$ &  & 4 \\
	$v_A$ [$\mathrm{km}\cdot\mathrm{s}^{-1}$] &  & 6 \\
	$z_h$ [kpc] &  & 5 \\
	\hline
	$\varepsilon$ &  & 0.01 \\
	$\delta_\parallel$ &  & 0.16 \\
	$\delta_\perp$ &  & 0.46 \\
	\enddata
\end{deluxetable}


\begin{table*}
	\begin{center}
		\begin{tabular}{|c|ccc|ccc|}
			\hline
			& \multicolumn{3}{c|}{Background} & \multicolumn{3}{c|}{Local source} \\
			\hline
			Element & A$^\dagger$ & ~~~$\nu$~~~  & ~~~$\mathcal R_{c}$~~~ & ~~~$q_0$~~~~~ & ~~~~~$\alpha$~~~ & ~~~${\cal R}'_c$~~~ \\
			\hline
			& $[({\rm m}^2\cdot {\rm sr}\cdot {\rm s}\cdot {\rm GeV})^{-1}]$ & & [PV] & [GeV$^{-1}$] & &  [TV] \\
			\hline
			p   & $4.41\times 10^{-2}$    & 2.36   &  7  & $3.20\times 10^{52}$  & 2.10 & 30 \\
			He & $1.18\times 10^{-2}$   & 2.30     &  7  & $1.10\times 10^{52}$  & 2.02  &  30  \\
			\hline
		\end{tabular}\\
		$^\dagger${The normalization is set at total energy $E = 100$ GeV.}
	\end{center}
	\caption{Fitted injection parameters of the background and local sources.}
	\label{tab:para_inj}
\end{table*}

We begin by calibrating the transport parameters of the SDP framework using the boron-to-carbon (B/C) ratio. The key quantities constrained in the fitting process include the diffusion normalization and slope $(D_0, \delta_0)$, the modulation factors $(N_m, \xi, n)$, the Alfv\'en velocity $v_A$, and the halo scale height $z_h$. The fitted B/C ratio is shown in Fig. \ref{fig:BCratio}, and the corresponding propagation parameter values are listed in Tab. \ref{tab:para_trans}. The injection spectra of background nuclei of species $i$ are characterized by the normalization $A_i$, spectral index $\nu_i$, and cutoff rigidity $R_c$. For the nearby component, the corresponding parameters are $(q_{0,i}, \alpha_i, R_c')$. The injection parameters are summarized in Tab. \ref{tab:para_inj}.
In the SDP model, high-energy CRs propagate to Earth through the inner halo, while low-energy CRs primarily diffuse in the outer halo. As a result, high-energy CRs arriving at Earth exhibit a harder spectrum, leading to the concave shape observed in the background proton and helium spectra. To account for the observed excesses of both proton and helium fluxes, the injection spectra from the local source must be harder than the averaged background. The cutoff rigidities of different compositions are regarded as the limits of acceleration in the sources and are assumed to be $Z$-dependent. To account for the softening at tens of TeV in proton and helium spectra, the cut-off rigidity of CRs from the Geminga SNR is $30$ TV.

To reproduce the spectral hardening observed around 200 GV, the nearby source is assigned a harder injection spectral index than the background component, and such a difference is physically expected. Observationally, the radio synchrotron spectral index $\gamma$ of shell-type SNRs typically falls within the approximate range $-0.8$ to $-0.2$, according to Green’s catalog \citep{2023ApJS..265...53R}. Using the standard relation between the radio index and the electron spectral index
$\alpha = -(\gamma-1)/2$, 
this corresponds to a cosmic-ray electron spectral index $\alpha$ between approximately $1.4$ and $2.6$. Furthermore, by fitting multiwavelength spectra of 35 SNRs with a one-zone model, \citet{2019ApJ...874...50Z} found that both the break energy and the low-energy spectral index (typically $\sim 1.3$--$2.3$) decrease with increasing SNR age, implying that younger remnants tend to produce softer spectra with higher break energies, whereas older remnants yield harder spectra with lower break energies. The value of $\alpha \sim 2.1$ inferred for Geminga here comfortably lies within this feasible range, while the softer background spectrum likely reflects the averaged contribution of numerous SNRs with diverse intrinsic spectral properties throughout the Galaxy.

\begin{figure*}[!ht]
	\includegraphics[width=0.5\textwidth]{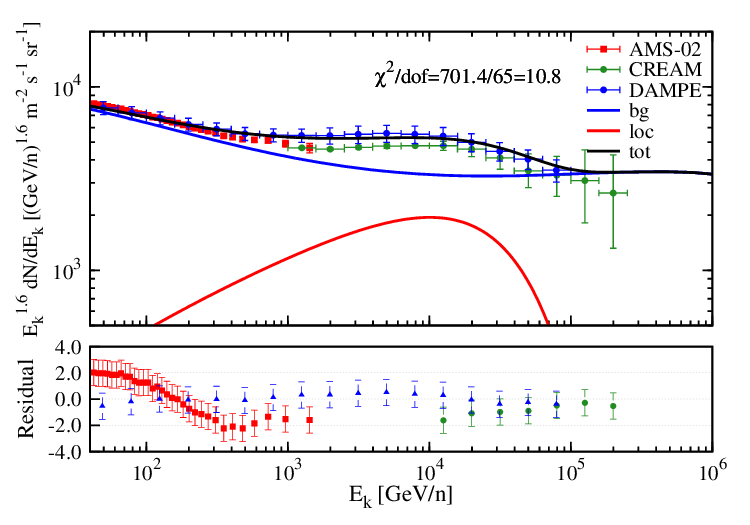}
	\includegraphics[width=0.5\textwidth]{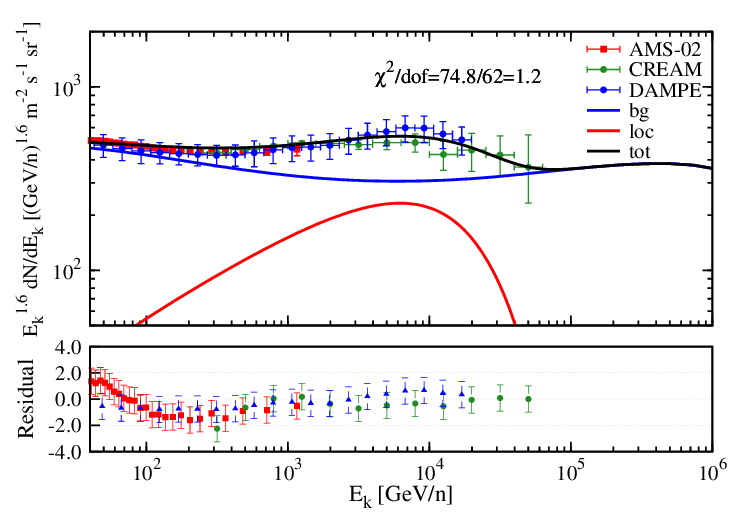}
	\caption{Top panels: modeled energy spectra of cosmic-ray proton (left) and helium nuclei (right). Measurements from DAMPE \citep{2019SciA....5.3793A, 2021PhRvL.126t1102A}, AMS-02 \citep{2015PhRvL.114q1103A, 2017PhRvL.119y1101A}, and CREAM-III \citep{2017ApJ...839....5Y} are shown for comparison. The blue curves illustrate the contribution from the large-scale Galactic background, while the red curves indicate the additional flux from a local source associated with the Geminga SNR. The black curves display the combined result of the two components. The corresponding reduced chi-squared ($\chi^2$/dof) values are also indicated in the panels. Bottom panels: the fitting residuals. The points of different colors represent the residuals with respect to the AMS-02(red), CREAM(green), and DAMPE(blue), respectively.
    }
    \label{fig:spec}
\end{figure*}

Fig.~\ref{fig:spec} shows the calculated proton and helium spectra together with the corresponding fit residuals\footnote{Residuals are defined as $(\Phi_{\rm exp}-\Phi_{\rm mod})/\sigma_{\rm exp}$, where $\Phi_{\rm exp}$ is the measured flux, $\Phi_{\rm mod}$ the model prediction, and $\sigma_{\rm exp}$ the experimental uncertainty.}. 
The top panels present the comparison between the model-predicted spectra and the observational data. To evaluate the quality of the fits, we compute the reduced chi-squared values, which are about 10.8 for protons and 1.2 for helium. The model reproduces the helium spectrum well, whereas the agreement for protons is relatively poorer. This discrepancy is largely driven by the deviation between the model prediction and the AMS-02 measurements. To illustrate this, the bottom panels display the fit residuals, which provide a direct visual measure of the agreement between the model and the data. For AMS-02, the residuals of the proton spectrum are significantly larger than those of helium, resulting in a higher reduced chi-squared value for protons. In contrast, the CREAM and DAMPE data extend to higher energies with larger uncertainties, leading to smaller normalized residuals that remain close to zero for both protons and helium. Nevertheless, the residuals of all three experiments fluctuate around zero within $3\sigma$, confirming that the model reproduces the observed spectra without significant systematic bias.

\begin{figure*}[!ht]
	\includegraphics[width=0.5\textwidth]{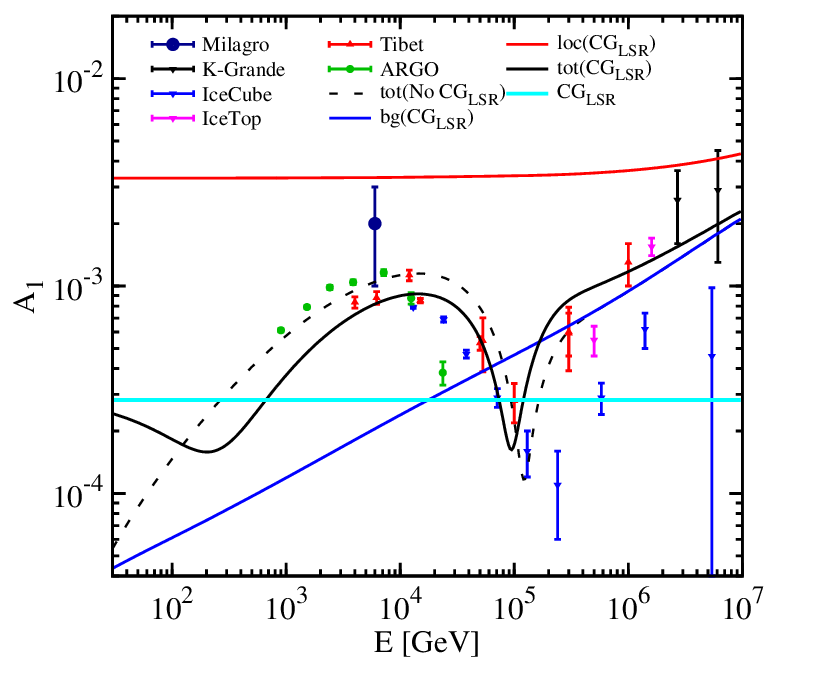}
	\includegraphics[width=0.5\textwidth]{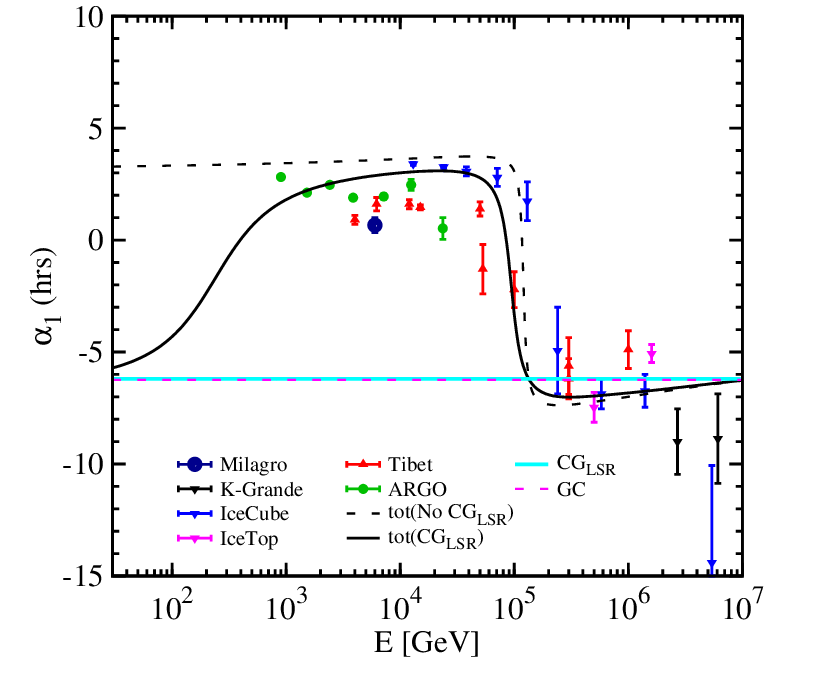}
	\caption{The model calculations of the amplitude (left) and phase (right) of the dipole anisotropies for the combined proton and helium flux. In the left panel, the blue solid line is the contribution from background sources, the red one is for Geminga SNR, and the cyan one is the result of Compton-Getting effect ($\rm CG_{LSR}$ effect). In the right panel, the cyan and magenta lines represent the directions from $\rm CG_{LSR}$ effect and the Galactic center respectively. The black solid and dashed lines in both panels are the total anisotropies with and without $\rm CG_{LSR}$ effect. Measurements are taken from Milagro (2009; \cite{2009ApJ...698.2121A}), K-Grande (2019; \cite{2019ApJ...870...91A}), IceCube (2010, 2012; \cite{2010ApJ...718L.194A, 2012ApJ...746...33A}), IceTop (2013; \cite{2013ApJ...765...55A}), ARGO-YBJ (2015; \cite{2015ApJ...809...90B}), Tibet (2005, 2017; \cite{2005ApJ...626L..29A, 2017ApJ...836..153A} respectively.  
	}
	\label{fig:aniso_cg}
\end{figure*}

In Fig.~\ref{fig:aniso_cg}, we compare the energy dependence of the dipole anisotropy with and without the CG effect, where the CG term corresponds to the $\rm CG_{LSR}$ contribution. The left panel shows the dipole amplitude: the blue and red solid lines correspond 
to the expected anisotropies from background sources and the Geminga SNR, respectively. In the SDP model, diffusion in the inner halo is significantly slower than in conventional models, resulting in a much smaller dipole amplitude from background sources below 100 TeV. However, the growth of anisotropy with energy is consistent with observations above 100 TeV. The anisotropy from the Geminga SNR is energy-independent due to time-dependent propagation effects. When the contributions are combined, a bump-like structure appears between 1 TeV and 100 TeV, caused by partial cancellation of CR streaming from background sources and the nearby Geminga SNR, as illustrated by the black dashed line. At around 100 TeV, a local minimum in the dipole amplitude is reached. Beyond this energy, the contribution from the Geminga SNR rapidly diminishes, and background sources dominate the anisotropy.

Regarding the dipole phase, below 100 TeV it points toward the direction of the local regular magnetic field 
(LRMF), as the CR Larmor radius is smaller than the LRMF scale, causing CRs to propagate preferentially along 
the field lines. Above 100 TeV, where the Larmor radius becomes comparable to the LRMF scale, diffusion becomes 
nearly isotropic, and the dipole phase gradually shifts toward the Galactic center.

When the CG effect is included, the anisotropy is further modified. The green solid line in Fig. \ref{fig:aniso_cg} represents the anisotropy from the $\rm CG_{LSR}$ effect alone. In the local standard of rest (LSR), 
the Sun's velocity is approximately 18 km/s, corresponding to a $\rm CG_{LSR}$ amplitude of about $3 \times 10^{-4}$, which is energy-independent in both amplitude and phase. The total anisotropy, shown by the black solid line, includes both CR propagation and the $\rm CG_{LSR}$ effect. Since the Sun's motion direction in the LSR is closer to the 
Galactic center, the $\rm CG_{LSR}$ effect suppresses the total dipole amplitude between 1 TeV and 100 TeV. More importantly, below $\sim$200 GeV, where anisotropies from both background and local sources diminish to $\sim10^{-4}$, the $\rm CG_{LSR}$ effect becomes more significant, causing the total anisotropy to increase again at lower energies. 


The $\rm CG_{LSR}$ effect also influences the dipole phase. Because the Sun’s motion in the LSR is oriented toward the 
Galactic center, the overall dipole phase is shifted down accordingly, as shown by the black solid line in the 
right panel of Fig. \ref{fig:aniso_cg}. Notably, as energy decreases from several TeV, the dipole phase 
gradually shifts toward the direction of $\rm CG_{LSR}$ effect. Quantitatively, the $\rm CG_{LSR}$ contribution modifies the dipole amplitude by approximately $30\%$–$50\%$ in the range $10^{2}$–$10^{3}\,\mathrm{GeV}$, and introduces a phase shift of a few hours in local sidereal time. 
Although the resulting change remains smaller than the current dispersion among experimental determinations, the predicted results—with and without the CG contribution—establish clear and testable benchmarks for ongoing and future ground-based observations. Building on recent LHAASO studies \citep{Liu:2025Pr}, the experiment is poised to explore cosmic-ray anisotropy down to energies of a few hundred~GeV, opening an exciting window into low-energy cosmic-ray behavior. While factors such as angular and energy resolution or solar modulation will need to be carefully accounted for, these challenges also present valuable opportunities to refine detection techniques and modeling frameworks. Successfully observing the predicted features would mark a significant milestone, providing a direct probe of the local cosmic-ray reference frame and deepening our understanding of cosmic-ray propagation in the heliosphere.



\begin{figure*}[!ht]
	\includegraphics[width=0.5\textwidth]{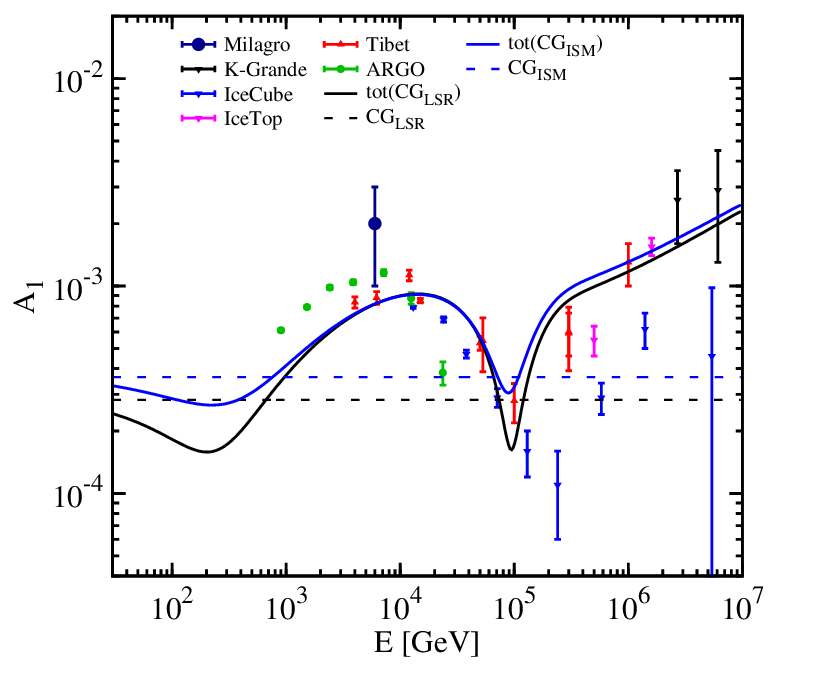}
	\includegraphics[width=0.5\textwidth]{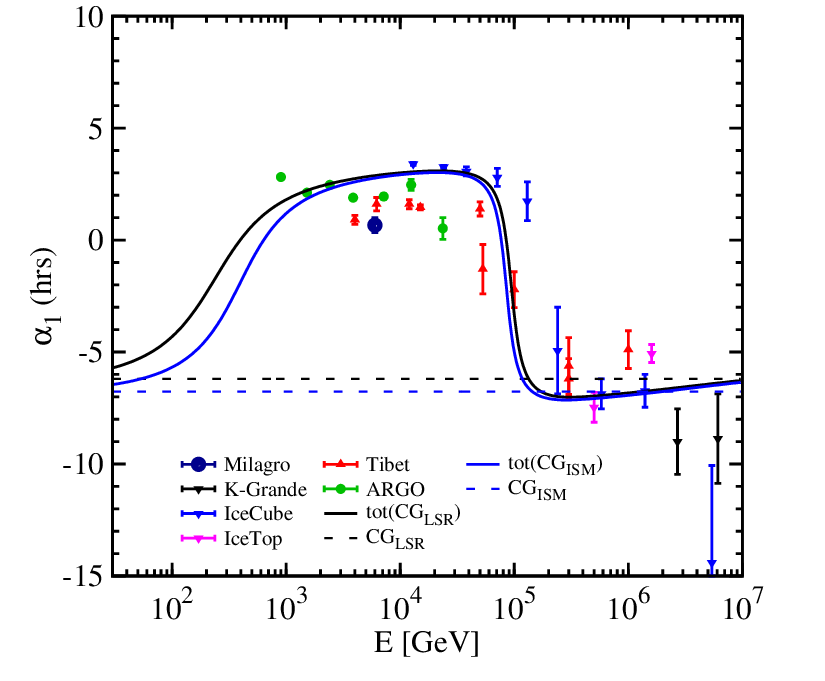}
	\caption{Comparison of predicted dipole anisotropy amplitude (left) and phase (right) under two assumptions for the solar motion: relative to the LSR (black) and relative to the local ISM (blue). Solid curves show the combined anisotropy, while dashed curves illustrate the individual CG contribution in each case.
    }
	\label{fig:aniso_cg2}
\end{figure*}

\begin{figure*}[!ht]
	\includegraphics[width=0.5\textwidth]{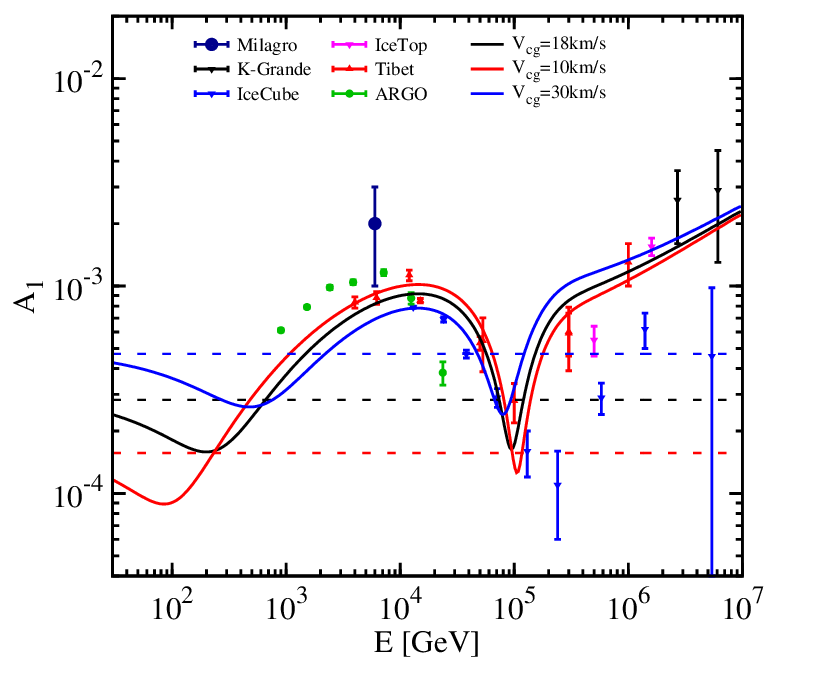}
	\includegraphics[width=0.5\textwidth]{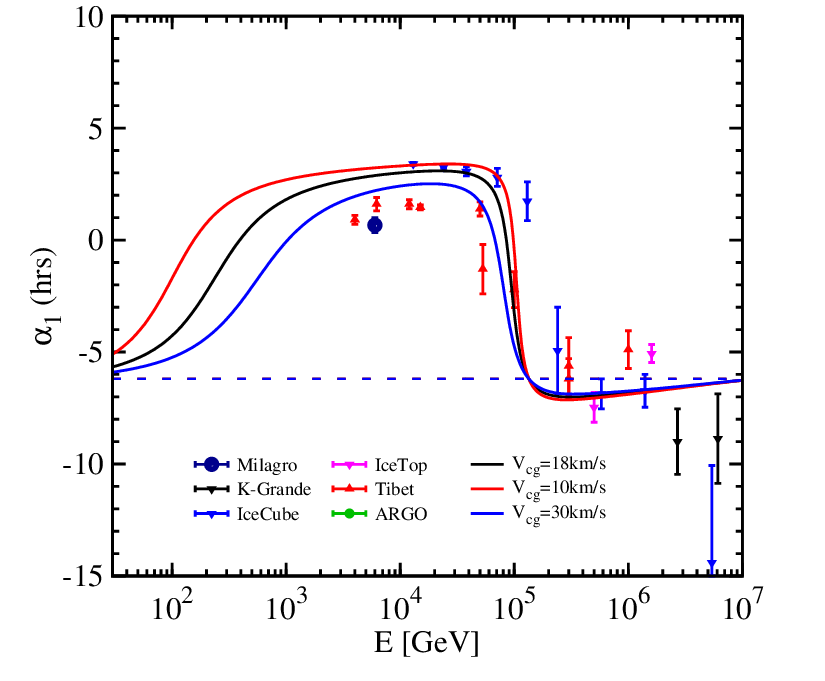}
	\caption{Predicted dipole anisotropy amplitude (left) and phase (right) for different assumed solar velocities with respect to the LSR, shown in red, black, and blue. Solid curves represent the total anisotropy including all contributions, while dashed curves display the corresponding $\rm CG_{LSR}$ terms for each velocity.
    }
	\label{fig:aniso_cg3}
\end{figure*}
 
However, current measurements of the Sun's local velocity still have uncertainties. Fig. \ref{fig:aniso_cg2} compares the CG effects in which the solar motion is relative to the LSR and the local ISM respectively, as mentioned above $\rm CG_{LSR}$ and $\rm CG_{ISM}$. Their impact on anisotropy is similar despite that they are different measurements, except that the expected total anisotropy below $100$ GeV is larger for the case when solar motion is relative to the local ISM due to the large velocity. Meanwhile, the flip of the dipole phase at lower energy in this case is faster. Fig. \ref{fig:aniso_cg3} illustrates the influence of different measurements of the Sun's velocity on the dipole anisotropy. As the velocity increases, the total amplitude reduces at TeV energies. However, below TeV, where the contribution from the density gradient of CRs is minimal, the $\rm CG_{LSR}$ effect can introduce substantial changes, when the velocity is large. Additionally, a higher velocity results in a more abrupt and earlier phase transition.

\begin{figure*}[!ht]
\includegraphics[width=0.5\textwidth]{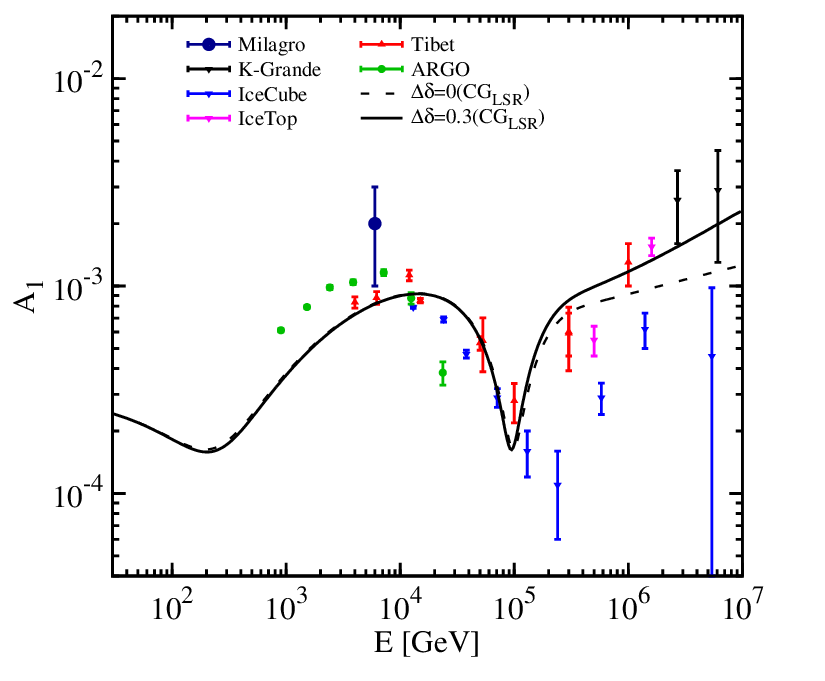}
\includegraphics[width=0.5\textwidth]{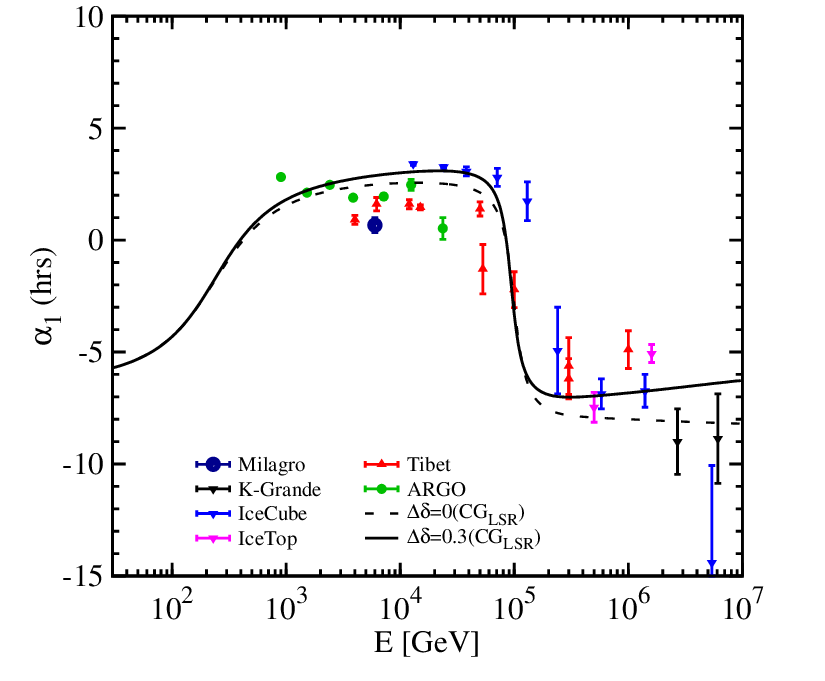}
\caption{Predicted amplitude (left) and phase (right) of the dipole anisotropy including the $\mathrm{CG}_{\mathrm{LSR}}$ contribution for two scenarios: $\Delta\delta = 0$ (dashed) and $\Delta\delta = 0.3$ (solid). The black solid line has the same meaning as in Fig.~\ref{fig:aniso_cg}, while the black dashed line represents the result for $\Delta\delta = 0$. 
}
\label{fig:aniso_cg4}
\end{figure*}

To illustrate the influence of $\delta_\perp$, defined as $\Delta\delta \equiv \delta_\perp-\delta_\parallel$, on the dipole anisotropy, we compute the amplitude and phase including the $\mathrm{CG}_{\mathrm{LSR}}$ contribution for two scenarios: $\Delta\delta = 0$ and $\Delta\delta = 0.3$, as shown in Fig.~\ref{fig:aniso_cg4}. At energies below $\sim 100$~TeV, the anisotropy is mainly governed by the nearby Geminga source, CRs preferentially propagate along the local regular magnetic field under strongly anisotropic diffusion ($D_\perp \ll D_\parallel$), so the difference between $\Delta\delta = 0$ and 0.3 is minor and both the amplitude and phase remain nearly identical. At higher energies, however, the large-scale Galactic background dominates. For $\Delta\delta = 0$, the transport remains highly anisotropic, suppressing the effective flux perpendicular to the magnetic field, which yields a smaller amplitude and a phase offset from the Galactic-center direction. In contrast, when $\Delta\delta = 0.3$, the perpendicular diffusion coefficient increases more rapidly with energy, driving the transport toward quasi-isotropy. Consequently, the amplitude rises and the phase gradually shifts toward the Galactic-center direction. This behavior is consistent with the standard anisotropic diffusion tensor picture and with previous studies showing that the energy dependence of $D_\perp$ can steer the observed dipole properties at high energies\citep{1999ApJ...520..204G}.

\section{Summary}
\label{sec:summary}

Currently, measurements of CR anisotropies are becoming increasingly precise. There is a growing recognition 
that large-scale dipole anisotropy carries valuable information about nearby sources and the local propagation environment. In our previous work, we proposed a self-consistent propagation scenario that simultaneously 
explains the spectral hardening and the energy-dependent anisotropy. 

In this study, we further investigate the impact of the Sun’s relative motion on the anisotropy, i.e. so-called Compton-Getting effect. The expected amplitude and phase from the CG effect is energy-independent. Previously, the CG effect is thought to originate from the Solar circular motion around the Galactic center with an average circular speed $v \simeq 220$ km/s. However, the experiments did not measure such large effect on the anisotropy. Thereupon, it is inferred that the Galactic CRs mostly co-rotate with the Galaxy. 

Here, we consider the Sun’s motion relative to the local standard rest or local plasma frame. We found that when considering this effect, the overall dipole amplitude decreases and the phase slightly deviates from the direction of the local regular magnetic field at tens of TeV energies. Below hundreds of GeV, when the anisotropy from the density gradient of CRs is small, the influence of the relative motion of the Sun becomes more prominent. Below $\sim 200$ GeV, the dipole amplitude even rises again and approaches the value of the $\rm CG_{LSR}$($\rm CG_{ISM}$) effect. Meanwhile, at several hundreds of GeV, there is another flip in the dipole phase, which points to the direction of the $\rm CG_{LSR}$($\rm CG_{ISM}$) effect. Quantitatively, the $\rm CG_{LSR}$ contribution alters the dipole amplitude by about $30\%$–$50\%$ at $10^{2}$–$10^{3}\,\mathrm{GeV}$ and induces a phase shift of a few hours in local sidereal time.
These features, shown with and without the $\rm CG_{LSR}$($\rm CG_{ISM}$) effect, 
could be confirmed by future anisotropy measurements in the 100 GeV to TeV energy range.

\section*{Acknowledgements}
WL \& YG accknowledge the support by the National Key R\&D program of China under the grant 2024YFA1611402 and the National Natural Science Foundation of China (12275279,12375103).

\bibliographystyle{apj}
\bibliography{ref1}

@article{Hou2025,
   author = {Chuanpeng Hou and Huirong Yan and Siqi Zhao and Parth Pavaskar},
   doi = {10.3847/2041-8213/AE0C97},
   issn = {2041-8205},
   issue = {2},
   journal = {The Astrophysical Journal Letters},
   month = {10},
   pages = {L28},
   title = {Energy Cascade and Damping in Fast-mode Compressible Turbulence},
   volume = {992},
   url = {https://iopscience.iop.org/article/10.3847/2041-8213/ae0c97},
   year = {2025}
}

@article{Zhao2025,
   author = {Siqi Zhao and Huirong Yan and Terry Z. Liu and Chuanpeng Hou},
   journal = {ApJ, submitted},
   month = {10},
   title = {Observations of the Relationship between Magnetic Anisotropy and Mode Composition in Low-$\beta$ Solar Wind Turbulence},
   url = {https://arxiv.org/pdf/2510.25636},
   year = {2025}
}

@ARTICLE{2022ApJ...926...94M,
       author = {{Maiti}, Snehanshu and {Makwana}, Kirit and {Zhang}, Heshou and {Yan}, Huirong},
        title = "{Cosmic-ray Transport in Magnetohydrodynamic Turbulence}",
      journal = {\apj},
         year = 2022,
        month = feb,
       volume = {926},
       number = {1},
          eid = {94},
        pages = {94},
          doi = {10.3847/1538-4357/ac46c8},
archivePrefix = {arXiv},
       eprint = {2108.01936},
 primaryClass = {astro-ph.HE},
       adsurl = {https://ui.adsabs.harvard.edu/abs/2022ApJ...926...94M}
}

@ARTICLE{2002PhRvL..89B1102Y,
       author = {{Yan}, Huirong and {Lazarian}, A.},
        title = "{Scattering of Cosmic Rays by Magnetohydrodynamic Interstellar Turbulence}",
      journal = {\prl},
         year = 2002,
        month = dec,
       volume = {89},
       number = {28},
          eid = {281102},
        pages = {281102},
          doi = {10.1103/PhysRevLett.89.281102},
archivePrefix = {arXiv},
       eprint = {astro-ph/0205285},
 primaryClass = {astro-ph},
       adsurl = {https://ui.adsabs.harvard.edu/abs/2002PhRvL..89B1102Y}
}

@ARTICLE{2004ApJ...614..757Y,
       author = {{Yan}, Huirong and {Lazarian}, A.},
        title = "{Cosmic-Ray Scattering and Streaming in Compressible Magnetohydrodynamic Turbulence}",
      journal = {\apj},
         year = 2004,
        month = oct,
       volume = {614},
       number = {2},
        pages = {757-769},
          doi = {10.1086/423733},
archivePrefix = {arXiv},
       eprint = {astro-ph/0408172},
 primaryClass = {astro-ph},
       adsurl = {https://ui.adsabs.harvard.edu/abs/2004ApJ...614..757Y}
}

@INPROCEEDINGS{Yan2022icrc,
       author = {{Yan}, H.},
        title = "{Magnetohydrodynamic turbulence and propagation of cosmic rays: theory confronted with observations}",
    booktitle = {37th International Cosmic Ray Conference},
         year = 2022,
        month = mar,
          eid = {38},
        pages = {38},
          doi = {10.22323/1.395.0038},
archivePrefix = {arXiv},
       eprint = {2109.07847},
 primaryClass = {astro-ph.HE},
       adsurl = {https://ui.adsabs.harvard.edu/abs/2022icrc.confE..38Y}
}

@ARTICLE{2017ApJ...842...54L,
       author = {{L{\'o}pez-Barquero}, V. and {Xu}, S. and {Desiati}, P. and {Lazarian}, A. and {Pogorelov}, N.~V. and {Yan}, H.},
        title = "{TeV Cosmic-Ray Anisotropy from the Magnetic Field at the Heliospheric Boundary}",
      journal = {\apj},
         year = 2017,
        month = jun,
       volume = {842},
       number = {1},
          eid = {54},
        pages = {54},
          doi = {10.3847/1538-4357/aa74d1},
archivePrefix = {arXiv},
       eprint = {1610.03097},
 primaryClass = {astro-ph.HE},
       adsurl = {https://ui.adsabs.harvard.edu/abs/2017ApJ...842...54L}
}

@BOOK{2002cra..book.....S,
       author = {{Schlickeiser}, Reinhard},
        title = "{Cosmic Ray Astrophysics}",
         year = 2002,
         publisher = {Springer},
       adsurl = {https://ui.adsabs.harvard.edu/abs/2002cra..book.....S}
}

@ARTICLE{2014ApJ...782...36E,
       author = {{Evoli}, Carmelo and {Yan}, Huirong},
        title = "{Cosmic Ray Propagation in Galactic Turbulence}",
      journal = {\apj},
         year = 2014,
        month = feb,
       volume = {782},
       number = {1},
          eid = {36},
        pages = {36},
          doi = {10.1088/0004-637X/782/1/36},
archivePrefix = {arXiv},
       eprint = {1310.5732},
 primaryClass = {astro-ph.HE},
       adsurl = {https://ui.adsabs.harvard.edu/abs/2014ApJ...782...36E}
}

@ARTICLE{2008ApJ...673..942Y,
       author = {{Yan}, Huirong and {Lazarian}, A.},
        title = "{Cosmic-Ray Propagation: Nonlinear Diffusion Parallel and Perpendicular to Mean Magnetic Field}",
      journal = {\apj},
         year = 2008,
        month = feb,
       volume = {673},
       number = {2},
        pages = {942-953},
          doi = {10.1086/524771},
archivePrefix = {arXiv},
       eprint = {0710.2617},
 primaryClass = {astro-ph},
       adsurl = {https://ui.adsabs.harvard.edu/abs/2008ApJ...673..942Y}
}

@ARTICLE{2006Sci...314..439A,
	author = {{Amenomori}, M. and {Ayabe}, S. and {Bi}, X.~J. and {Chen}, D. and {Cui}, S.~W. and {Danzengluobu} and {Ding}, L.~K. and {Ding}, X.~H. and {Feng}, C.~F. and {Feng}, Zhaoyang and {Feng}, Z.~Y. and {Gao}, X.~Y. and {Geng}, Q.~X. and {Guo}, H.~W. and {He}, H.~H. and {He}, M. and {Hibino}, K. and {Hotta}, N. and {Hu}, Haibing and {Hu}, H.~B. and {Huang}, J. and {Huang}, Q. and {Jia}, H.~Y. and {Kajino}, F. and {Kasahara}, K. and {Katayose}, Y. and {Kato}, C. and {Kawata}, K. and {Labaciren} and {Le}, G.~M. and {Li}, A.~F. and {Li}, J.~Y. and {Lou}, Y. -Q. and {Lu}, H. and {Lu}, S.~L. and {Meng}, X.~R. and {Mizutani}, K. and {Mu}, J. and {Munakata}, K. and {Nagai}, A. and {Nanjo}, H. and {Nishizawa}, M. and {Ohnishi}, M. and {Ohta}, I. and {Onuma}, H. and {Ouchi}, T. and {Ozawa}, S. and {Ren}, J.~R. and {Saito}, T. and {Saito}, T.~Y. and {Sakata}, M. and {Sako}, T.~K. and {Sasaki}, T. and {Shibata}, M. and {Shiomi}, A. and {Shirai}, T. and {Sugimoto}, H. and {Takita}, M. and {Tan}, Y.~H. and {Tateyama}, N. and {Torii}, S. and {Tsuchiya}, H. and {Udo}, S. and {Wang}, B. and {Wang}, H. and {Wang}, X. and {Wang}, Y.~G. and {Wu}, H.~R. and {Xue}, L. and {Yamamoto}, Y. and {Yan}, C.~T. and {Yang}, X.~C. and {Yasue}, S. and {Ye}, Z.~H. and {Yu}, G.~C. and {Yuan}, A.~F. and {Yuda}, T. and {Zhang}, H.~M. and {Zhang}, J.~L. and {Zhang}, N.~J. and {Zhang}, X.~Y. and {Zhang}, Y. and {Zhang}, Yi and {Zhaxisangzhu} and {Zhou}, X.~X. and {Tibet AS{\ensuremath{\gamma}} Collaboration}},
	title = "{Anisotropy and Corotation of Galactic Cosmic Rays}",
	journal = {Science},
	year = 2006,
	month = oct,
	volume = {314},
	number = {5798},
	pages = {439-443},
	doi = {10.1126/science.1131702},
	archivePrefix = {arXiv},
	eprint = {astro-ph/0610671},
	primaryClass = {astro-ph},
	adsurl = {https://ui.adsabs.harvard.edu/abs/2006Sci...314..439A}
}

@ARTICLE{2017ApJ...836..153A,
       author = {{Amenomori}, M. and {Bi}, X.~J. and {Chen}, D. and {Chen}, T.~L. and {Chen}, W.~Y. and {Cui}, S.~W. and {Danzengluobu} and {Ding}, L.~K. and {Feng}, C.~F. and {Feng}, Zhaoyang and {Feng}, Z.~Y. and {Gou}, Q.~B. and {Guo}, Y.~Q. and {He}, H.~H. and {He}, Z.~T. and {Hibino}, K. and {Hotta}, N. and {Hu}, Haibing and {Hu}, H.~B. and {Huang}, J. and {Jia}, H.~Y. and {Jiang}, L. and {Kajino}, F. and {Kasahara}, K. and {Katayose}, Y. and {Kato}, C. and {Kawata}, K. and {Kozai}, M. and {Labaciren} and {Le}, G.~M. and {Li}, A.~F. and {Li}, H.~J. and {Li}, W.~J. and {Liu}, C. and {Liu}, J.~S. and {Liu}, M.~Y. and {Lu}, H. and {Meng}, X.~R. and {Miyazaki}, T. and {Mizutani}, K. and {Munakata}, K. and {Nakajima}, T. and {Nakamura}, Y. and {Nanjo}, H. and {Nishizawa}, M. and {Niwa}, T. and {Ohnishi}, M. and {Ohta}, I. and {Ozawa}, S. and {Qian}, X.~L. and {Qu}, X.~B. and {Saito}, T. and {Saito}, T.~Y. and {Sakata}, M. and {Sako}, T.~K. and {Shao}, J. and {Shibata}, M. and {Shiomi}, A. and {Shirai}, T. and {Sugimoto}, H. and {Takita}, M. and {Tan}, Y.~H. and {Tateyama}, N. and {Torii}, S. and {Tsuchiya}, H. and {Udo}, S. and {Wang}, H. and {Wu}, H.~R. and {Xue}, L. and {Yamamoto}, Y. and {Yamauchi}, K. and {Yang}, Z. and {Yuan}, A.~F. and {Yuda}, T. and {Zhai}, L.~M. and {Zhang}, H.~M. and {Zhang}, J.~L. and {Zhang}, X.~Y. and {Zhang}, Y. and {Zhang}, Yi and {Zhang}, Ying and {Zhaxisangzhu} and {Zhou}, X.~X. and {Tibet AS{\ensuremath{\gamma}} Collaboration}},
        title = "{Northern Sky Galactic Cosmic Ray Anisotropy between 10 and 1000 TeV with the Tibet Air Shower Array}",
      journal = {\apj},
         year = 2017,
        month = feb,
       volume = {836},
       number = {2},
          eid = {153},
        pages = {153},
          doi = {10.3847/1538-4357/836/2/153},
archivePrefix = {arXiv},
       eprint = {1701.07144},
 primaryClass = {astro-ph.HE},
       adsurl = {https://ui.adsabs.harvard.edu/abs/2017ApJ...836..153A}
}

@ARTICLE{2024NuScT..35..149P,
       author = {{Pan}, Xu and {Jiang}, Wei and {Yue}, Chuan and {Lei}, Shi-Jun and {Cui}, Yu-Xin and {Yuan}, Qiang},
        title = "{Simulation study of the performance of the Very Large Area gamma-ray Space Telescope}",
      journal = {Nuclear Science and Techniques},
         year = 2024,
        month = sep,
       volume = {35},
       number = {9},
          eid = {149},
        pages = {149},
          doi = {10.1007/s41365-024-01499-x},
archivePrefix = {arXiv},
       eprint = {2407.16973},
 primaryClass = {astro-ph.IM},
       adsurl = {https://ui.adsabs.harvard.edu/abs/2024NuScT..35..149P}
}

@ARTICLE{2019arXiv190502773C,
       author = {{Cao}, Zhen and {della Volpe}, D. and {Liu}, Siming and {Editors} and {:} and {Bi}, Xiaojun and {Chen}, Yang and {D'Ettorre Piazzoli}, B. and {Feng}, Li and {Jia}, Huanyu and {Li}, Zhuo and {Ma}, Xinhua and {Wang}, Xiangyu and {Zhang}, Xiao and {Referees}, External and {:} and {Qie}, Xiushu and {Hu}, Hongbo and {Referees}, Internal and {:} and {S{\'a}iz}, Alejandro and {Yang}, Ruizhi and {Contributors} and {:} and {Addazi}, Andrea and {Belotsky}, Konstantin and {Beylin}, Vitaly and {Bi}, Yu-Jiang and {Che}, Ming-Jun and {Chen}, Song-Zhan and {Cheng}, Yao-Dong and {Chiavassa}, Andrea and {Cirelli}, Marco and {Di Sciascio}, Giuseppe and {Esmaili}, Arman and {Fang}, Kun and {Fornengo}, Nicolao and {Gou}, Quanbu and {Guo}, Yi-Qing and {Gan}, Qingyu and {Gong}, Guang-Hua and {Gu}, Min-Hao and {He}, Haoning and {He}, Hui-Hai and {Hou}, Chao and {Huang}, Xing-Tao and {Huang}, Wen-Hao and {Kachekriess}, Michael and {Khlopov}, Maxim and {Korchagin}, Vladimir and {Korochkin}, Alexander and {Kuksa}, Vladimir and {Ksenofontov}, Leonid T. and {Liu}, Ye and {Liu}, Ruo-Yu and {Liu}, Cheng and {Marciano}, Antonino and {Martineau-Huynh}, Olivier and {Martraire}, Diane and {Ma}, Lingling and {Neronov}, Andrii and {Panci}, Paolo and {Pasechnick}, Roman and {Ruffolo}, David and {Sakharov}, Alexander and {Sala}, Filippo and {Semikoz}, Dimiri and {Shchegolev}, Oleg and {Serpico}, Pasquale Dario and {Sheng}, Xiang-Dong and {Stenkin}, Yuri V. and {Tam}, P.~H. Thomas and {Vernetto}, Silvia and {Vallania}, Piero and {Volchanskiy}, Nikolay and {Wang}, Zhongxiang and {Wang}, Kai and {Wang}, Xiang-Yu and {Wu}, Han-Rong and {Wu}, Chao-Yong and {Wu}, Sha and {Xiao}, Gang and {Yang}, Rui-zhi and {Yan}, Dahai and {Yao}, Zhi-Guo and {Yin}, Pengfei and {Yuan}, Qiang and {Zhang}, Xiao and {Zeng}, Houdun and {Zhang}, Shou-Shan and {Zhang}, Yi and {Zhou}, Xunxiu and {Zhu}, Hui and {Zuo}, Xiong},
        title = "{The Large High Altitude Air Shower Observatory (LHAASO) Science Book (2021 Edition)}",
      journal = {arXiv e-prints},
         year = 2019,
        month = may,
          eid = {arXiv:1905.02773},
        pages = {arXiv:1905.02773},
          doi = {10.48550/arXiv.1905.02773},
archivePrefix = {arXiv},
       eprint = {1905.02773},
 primaryClass = {astro-ph.HE},
       adsurl = {https://ui.adsabs.harvard.edu/abs/2019arXiv190502773C}
}

@ARTICLE{2022SciBu..67.2162D,
       author = {{Dampe Collaboration}},
        title = "{Detection of spectral hardenings in cosmic-ray boron-to-carbon and boron-to-oxygen flux ratios with DAMPE}",
      journal = {Science Bulletin},
         year = 2022,
        month = nov,
       volume = {67},
       number = {21},
        pages = {2162-2166},
          doi = {10.1016/j.scib.2022.10.002},
archivePrefix = {arXiv},
       eprint = {2210.08833},
 primaryClass = {astro-ph.HE},
       adsurl = {https://ui.adsabs.harvard.edu/abs/2022SciBu..67.2162D}
}

@ARTICLE{2007PhRvD..75f2003G,
	author = {{Guillian}, G. and {Hosaka}, J. and {Ishihara}, K. and {Kameda}, J. and 
	{Koshio}, Y. and {Minamino}, A. and {Mitsuda}, C. and {Miura}, M. and 
	{Moriyama}, S. and {Nakahata}, M. and {Namba}, T. and {Obayashi}, Y. and 
	{Ogawa}, H. and {Shiozawa}, M. and {Suzuki}, Y. and {Takeda}, A. and 
	{Takeuchi}, Y. and {Yamada}, S. and {Higuchi}, I. and {Ishitsuka}, M. and 
	{Kajita}, T. and {Kaneyuki}, K. and {Mitsuka}, G. and {Nakayama}, S. and 
	{Nishino}, H. and {Okada}, A. and {Okumura}, K. and {Saji}, C. and 
	{Takenaga}, Y. and {Desai}, S. and {Kearns}, E. and {Stone}, J.~L. and 
	{Sulak}, L.~R. and {Wang}, W. and {Goldhaber}, M. and {Casper}, D. and 
	{Gajewski}, W. and {Griskevich}, J. and {Kropp}, W.~R. and {Liu}, D.~W. and 
	{Mine}, S. and {Smy}, M.~B. and {Sobel}, H.~W. and {Vagins}, M.~R. and 
	{Ganezer}, K.~S. and {Hill}, J. and {Keig}, W.~E. and {Scholberg}, K. and 
	{Walter}, C.~W. and {Ellsworth}, R.~W. and {Tasaka}, S. and 
	{Kibayashi}, A. and {Learned}, J.~G. and {Matsuno}, S. and {Messier}, M.~D. and 
	{Hayato}, Y. and {Ichikawa}, A.~K. and {Ishida}, T. and {Ishii}, T. and 
	{Iwashita}, T. and {Kobayashi}, T. and {Nakadaira}, T. and {Nakamura}, K. and 
	{Nitta}, K. and {Oyama}, Y. and {Totsuka}, Y. and {Suzuki}, A.~T. and 
	{Hasegawa}, M. and {Kato}, I. and {Maesaka}, H. and {Nakaya}, T. and 
	{Nishikawa}, K. and {Sato}, H. and {Yamamoto}, S. and {Yokoyama}, M. and 
	{Haines}, T.~J. and {Dazeley}, S. and {Hatakeyama}, S. and {Svoboda}, R. and 
	{Blaufuss}, E. and {Goodman}, J.~A. and {Sullivan}, G.~W. and 
	{Turcan}, D. and {Habig}, A. and {Fukuda}, Y. and {Itow}, Y. and 
	{Sakuda}, M. and {Yoshida}, M. and {Kim}, S.~B. and {Yoo}, J. and 
	{Okazawa}, H. and {Ishizuka}, T. and {Jung}, C.~K. and {Kato}, T. and 
	{Kobayashi}, K. and {Malek}, M. and {Mauger}, C. and {McGrew}, C. and 
	{Sharkey}, E. and {Yanagisawa}, C. and {Gando}, Y. and {Hasegawa}, T. and 
	{Inoue}, K. and {Shirai}, J. and {Suzuki}, A. and {Nishijima}, K. and 
	{Ishino}, H. and {Watanabe}, Y. and {Koshiba}, M. and {Kielczewska}, D. and 
	{Berns}, H.~G. and {Gran}, R. and {Shiraishi}, K.~K. and {Stachyra}, A.~L. and 
	{Washburn}, K. and {Wilkes}, R.~J. and {Munakata}, K.},
	title = "{Observation of the anisotropy of 10TeV primary cosmic ray nuclei flux with the Super-Kamiokande-I detector}",
	journal = {\prd},
	eprint = {astro-ph/0508468},
	year = 2007,
	month = mar,
	volume = 75,
	number = 6,
	eid = {062003},
	pages = {062003},
	doi = {10.1103/PhysRevD.75.062003},
	adsurl = {http://adsabs.harvard.edu/abs/2007PhRvD..75f2003G}
}

@ARTICLE{2008PhRvL.101v1101A,
	author = {{Abdo}, A.~A. and {Allen}, B. and {Aune}, T. and {Berley}, D. and 
	{Blaufuss}, E. and {Casanova}, S. and {Chen}, C. and {Dingus}, B.~L. and 
	{Ellsworth}, R.~W. and {Fleysher}, L. and {Fleysher}, R. and 
	{Gonzalez}, M.~M. and {Goodman}, J.~A. and {Hoffman}, C.~M. and 
	{H{\"u}ntemeyer}, P.~H. and {Kolterman}, B.~E. and {Lansdell}, C.~P. and 
	{Linnemann}, J.~T. and {McEnery}, J.~E. and {Mincer}, A.~I. and 
	{Nemethy}, P. and {Noyes}, D. and {Pretz}, J. and {Ryan}, J.~M. and 
	{Parkinson}, P.~M.~S. and {Shoup}, A. and {Sinnis}, G. and {Smith}, A.~J. and 
	{Sullivan}, G.~W. and {Vasileiou}, V. and {Walker}, G.~P. and 
	{Williams}, D.~A. and {Yodh}, G.~B.},
	title = "{Discovery of Localized Regions of Excess 10-TeV Cosmic Rays}",
	journal = {Physical Review Letters},
	archivePrefix = "arXiv",
	eprint = {0801.3827},
	year = 2008,
	month = nov,
	volume = 101,
	number = 22,
	eid = {221101},
	pages = {221101},
	doi = {10.1103/PhysRevLett.101.221101},
	adsurl = {http://adsabs.harvard.edu/abs/2008PhRvL.101v1101A}
}

@ARTICLE{2009ApJ...698.2121A,
	author = {{Abdo}, A.~A. and {Allen}, B.~T. and {Aune}, T. and {Berley}, D. and 
	{Casanova}, S. and {Chen}, C. and {Dingus}, B.~L. and {Ellsworth}, R.~W. and 
	{Fleysher}, L. and {Fleysher}, R. and {Gonzalez}, M.~M. and 
	{Goodman}, J.~A. and {Hoffman}, C.~M. and {Hopper}, B. and {H{\"u}ntemeyer}, P.~H. and 
	{Kolterman}, B.~E. and {Lansdell}, C.~P. and {Linnemann}, J.~T. and 
	{McEnery}, J.~E. and {Mincer}, A.~I. and {Nemethy}, P. and {Noyes}, D. and 
	{Pretz}, J. and {Ryan}, J.~M. and {Parkinson}, P.~M.~S. and 
	{Shoup}, A. and {Sinnis}, G. and {Smith}, A.~J. and {Sullivan}, G.~W. and 
	{Vasileiou}, V. and {Walker}, G.~P. and {Williams}, D.~A. and 
	{Yodh}, G.~B.},
	title = "{The Large-Scale Cosmic-Ray Anisotropy as Observed with Milagro}",
	journal = {apj},
	archivePrefix = "arXiv",
	eprint = {0806.2293},
	year = 2009,
	month = jun,
	volume = 698,
	pages = {2121-2130},
	doi = {10.1088/0004-637X/698/2/2121},
	adsurl = {http://adsabs.harvard.edu/abs/2009ApJ...698.2121A}
}

@ARTICLE{2010ApJ...718L.194A,
	author = {{Abbasi}, R. and {Abdou}, Y. and {Abu-Zayyad}, T. and {Adams}, J. and 
	{Aguilar}, J.~A. and {Ahlers}, M. and {Andeen}, K. and {Auffenberg}, J. and 
	{Bai}, X. and {Baker}, M. and et al.},
	title = "{Measurement of the Anisotropy of Cosmic-ray Arrival Directions with IceCube}",
	journal = {apj},
	archivePrefix = "arXiv",
	eprint = {1005.2960},
	primaryClass = "astro-ph.HE",
	year = 2010,
	month = aug,
	volume = 718,
	pages = {L194-L198},
	doi = {10.1088/2041-8205/718/2/L194},
	adsurl = {http://adsabs.harvard.edu/abs/2010ApJ...718L.194A}
}

@ARTICLE{2011ApJ...740...16A,
	author = {{Abbasi}, R. and {Abdou}, Y. and {Abu-Zayyad}, T. and {Adams}, J. and 
	{Aguilar}, J.~A. and {Ahlers}, M. and {Altmann}, D. and {Andeen}, K. and 
	{Auffenberg}, J. and {Bai}, X. and et al.},
	title = "{Observation of Anisotropy in the Arrival Directions of Galactic Cosmic Rays at Multiple Angular Scales with IceCube}",
	journal = {apj},
	archivePrefix = "arXiv",
	eprint = {1105.2326},
	primaryClass = "astro-ph.HE",
	year = 2011,
	month = oct,
	volume = 740,
	eid = {16},
	pages = {16},
	doi = {10.1088/0004-637X/740/1/16},
	adsurl = {http://adsabs.harvard.edu/abs/2011ApJ...740...16A}
}

@ARTICLE{2012ApJ...746...33A,
	author = {{Abbasi}, R. and {Abdou}, Y. and {Abu-Zayyad}, T. and {Ackermann}, M. and 
	{Adams}, J. and {Aguilar}, J.~A. and {Ahlers}, M. and {Allen}, M.~M. and 
	{Altmann}, D. and {Andeen}, K. and et al.},
	title = "{Observation of Anisotropy in the Galactic Cosmic-Ray Arrival Directions at 400 TeV with IceCube}",
	journal = {apj},
	archivePrefix = "arXiv",
	eprint = {1109.1017},
	primaryClass = "hep-ex",
	year = 2012,
	month = feb,
	volume = 746,
	eid = {33},
	pages = {33},
	doi = {10.1088/0004-637X/746/1/33},
	adsurl = {http://adsabs.harvard.edu/abs/2012ApJ...746...33A}
}

@ARTICLE{2013ApJ...765...55A,
	author = {{Aartsen}, M.~G. and {Abbasi}, R. and {Abdou}, Y. and {Ackermann}, M. and 
	{Adams}, J. and {Aguilar}, J.~A. and {Ahlers}, M. and {Altmann}, D. and 
	{Andeen}, K. and {Auffenberg}, J. and et al.},
	title = "{Observation of Cosmic-Ray Anisotropy with the IceTop Air Shower Array}",
	journal = {apj},
	archivePrefix = "arXiv",
	eprint = {1210.5278},
	primaryClass = "astro-ph.HE",
	year = 2013,
	month = mar,
	volume = 765,
	eid = {55},
	pages = {55},
	doi = {10.1088/0004-637X/765/1/55},
	adsurl = {http://adsabs.harvard.edu/abs/2013ApJ...765...55A}
}

@ARTICLE{2016ApJ...826..220A,
	author = {{Aartsen}, M.~G. and {Abraham}, K. and {Ackermann}, M. and {Adams}, J. and 
	{Aguilar}, J.~A. and {Ahlers}, M. and {Ahrens}, M. and {Altmann}, D. and 
	{Anderson}, T. and {Ansseau}, I. and et al.},
	title = "{Anisotropy in Cosmic-Ray Arrival Directions in the Southern Hemisphere Based on Six Years of Data from the IceCube Detector}",
	journal = {apj},
	archivePrefix = "arXiv",
	eprint = {1603.01227},
	primaryClass = "astro-ph.HE",
	year = 2016,
	month = aug,
	volume = 826,
	eid = {220},
	pages = {220},
	doi = {10.3847/0004-637X/826/2/220},
	adsurl = {http://adsabs.harvard.edu/abs/2016ApJ...826..220A}
}

@ARTICLE{2013PhRvD..88h2001B,
	author = {{Bartoli}, B. and {Bernardini}, P. and {Bi}, X.~J. and {Bolognino}, I. and 
	{Branchini}, P. and {Budano}, A. and {Calabrese Melcarne}, A.~K. and 
	{Camarri}, P. and {Cao}, Z. and {Cardarelli}, R. and {Catalanotti}, S. and 
	{Chen}, S.~Z. and {Chen}, T.~L. and {Creti}, P. and {Cui}, S.~W. and 
	{Dai}, B.~Z. and {D'Amone}, A. and {Danzengluobu} and {De Mitri}, I. and 
	{D'Ettorre Piazzoli}, B. and {Di Girolamo}, T. and {Di Sciascio}, G. and 
	{Feng}, C.~F. and {Feng}, Z. and {Feng}, Z. and {Gou}, Q.~B. and 
	{Guo}, Y.~Q. and {He}, H.~H. and {Hu}, H. and {Hu}, H. and {Iacovacci}, M. and 
	{Iuppa}, R. and {Jia}, H.~Y. and {Labaciren} and {Li}, H.~J. and 
	{Liguori}, G. and {Liu}, C. and {Liu}, J. and {Liu}, M.~Y. and 
	{Lu}, H. and {Ma}, X.~H. and {Mancarella}, G. and {Mari}, S.~M. and 
	{Marsella}, G. and {Martello}, D. and {Mastroianni}, S. and 
	{Montini}, P. and {Ning}, C.~C. and {Panareo}, M. and {Panico}, B. and 
	{Perrone}, L. and {Pistilli}, P. and {Ruggieri}, F. and {Salvini}, P. and 
	{Santonico}, R. and {Sbano}, S.~N. and {Shen}, P.~R. and {Sheng}, X.~D. and 
	{Shi}, F. and {Surdo}, A. and {Tan}, Y.~H. and {Vallania}, P. and 
	{Vernetto}, S. and {Vigorito}, C. and {Wang}, H. and {Wu}, C.~Y. and 
	{Wu}, H.~R. and {Xue}, L. and {Yan}, Y.~X. and {Yang}, Q.~Y. and 
	{Yang}, X.~C. and {Yao}, Z.~G. and {Yuan}, A.~F. and {Zha}, M. and 
	{Zhang}, H.~M. and {Zhang}, L. and {Zhang}, X.~Y. and {Zhang}, Y. and 
	{Zhaxiciren} and {Zhaxisangzhu} and {Zhou}, X.~X. and {Zhu}, F.~R. and 
	{Zhu}, Q.~Q. and {Zizzi}, G.},
	title = "{Medium scale anisotropy in the TeV cosmic ray flux observed by ARGO-YBJ}",
	journal = {prd},
	archivePrefix = "arXiv",
	eprint = {1309.6182},
	primaryClass = "astro-ph.HE",
	year = 2013,
	month = oct,
	volume = 88,
	number = 8,
	eid = {082001},
	pages = {082001},
	doi = {10.1103/PhysRevD.88.082001},
	adsurl = {http://adsabs.harvard.edu/abs/2013PhRvD..88h2001B}
}

@ARTICLE{2014ApJ...796..108A,
	author = {{Abeysekara}, A.~U. and {Alfaro}, R. and {Alvarez}, C. and {{\'A}lvarez}, J.~D. and 
	{Arceo}, R. and {Arteaga-Vel{\'a}zquez}, J.~C. and {Ayala Solares}, H.~A. and 
	{Barber}, A.~S. and {Baughman}, B.~M. and {Bautista-Elivar}, N. and 
	{Belmont}, E. and {BenZvi}, S.~Y. and {Berley}, D. and {Bonilla Rosales}, M. and 
	{Braun}, J. and {Caballero-Mora}, K.~S. and {Carrami{\~n}ana}, A. and 
	{Castillo}, M. and {Cotti}, U. and {Cotzomi}, J. and {de la Fuente}, E. and 
	{De Le{\'o}n}, C. and {DeYoung}, T. and {Diaz Hernandez}, R. and 
	{D{\'{\i}}az-V{\'e}lez}, J.~C. and {Dingus}, B.~L. and {DuVernois}, M.~A. and 
	{Ellsworth}, R.~W. and {Fiorino}, D.~W. and {Fraija}, N. and 
	{Galindo}, A. and {Garfias}, F. and {Gonz{\'a}lez}, M.~M. and 
	{Goodman}, J.~A. and {Gussert}, M. and {Hampel-Arias}, Z. and 
	{Harding}, J.~P. and {H{\"u}ntemeyer}, P. and {Hui}, C.~M. and 
	{Imran}, A. and {Iriarte}, A. and {Karn}, P. and {Kieda}, D. and 
	{Kunde}, G.~J. and {Lara}, A. and {Lauer}, R.~J. and {Lee}, W.~H. and 
	{Lennarz}, D. and {Le{\'o}n Vargas}, H. and {Linnemann}, J.~T. and 
	{Longo}, M. and {Luna-Garc{\'{\i}}a}, R. and {Malone}, K. and 
	{Marinelli}, A. and {Marinelli}, S.~S. and {Martinez}, H. and 
	{Martinez}, O. and {Mart{\'{\i}}nez-Castro}, J. and {Matthews}, J.~A.~J. and 
	{McEnery}, J. and {Mendoza Torres}, E. and {Miranda-Romagnoli}, P. and 
	{Moreno}, E. and {Mostaf{\'a}}, M. and {Nellen}, L. and {Newbold}, M. and 
	{Noriega-Papaqui}, R. and {Oceguera-Becerra}, T. and {Patricelli}, B. and 
	{Pelayo}, R. and {P{\'e}rez-P{\'e}rez}, E.~G. and {Pretz}, J. and 
	{Rivi{\`e}re}, C. and {Rosa-Gonz{\'a}lez}, D. and {Ruiz-Velasco}, E. and 
	{Ryan}, J. and {Salazar}, H. and {Salesa Greus}, F. and {Sandoval}, A. and 
	{Schneider}, M. and {Sinnis}, G. and {Smith}, A.~J. and {Sparks Woodle}, K. and 
	{Springer}, R.~W. and {Taboada}, I. and {Toale}, P.~A. and {Tollefson}, K. and 
	{Torres}, I. and {Ukwatta}, T.~N. and {Villase{\~n}or}, L. and 
	{Weisgarber}, T. and {Westerhoff}, S. and {Wisher}, I.~G. and 
	{Wood}, J. and {Yodh}, G.~B. and {Younk}, P.~W. and {Zaborov}, D. and 
	{Zepeda}, A. and {Zhou}, H. and {HAWC Collaboration}},
	title = "{Observation of Small-scale Anisotropy in the Arrival Direction Distribution of TeV Cosmic Rays with HAWC}",
	journal = {apj},
	archivePrefix = "arXiv",
	eprint = {1408.4805},
	primaryClass = "astro-ph.HE",
	year = 2014,
	month = dec,
	volume = 796,
	eid = {108},
	pages = {108},
	doi = {10.1088/0004-637X/796/2/108},
	adsurl = {http://adsabs.harvard.edu/abs/2014ApJ...796..108A}
}

@Article{2012JCAP...01..011B,
	Title                    = {{Diffusive propagation of cosmic rays from supernova remnants in the Galaxy. II: anisotropy}},
	Author                   = {{Blasi}, P. and {Amato}, E.},
	Journal                  = {jcap},
	Year                     = {2012},
	
	Month                    = jan,
	Pages                    = {11},
	Volume                   = {1},
	Adsurl                   = {http://adsabs.harvard.edu/abs/2012JCAP...01..011B},
	Archiveprefix            = {arXiv},
	Doi                      = {10.1088/1475-7516/2012/01/011},
	Eid                      = {011},
	Eprint                   = {1105.4529},
	Primaryclass             = {astro-ph.HE}
}

@ARTICLE{2017PhRvD..96b3006L,
	author = {{Liu}, Wei and {Bi}, Xiao-Jun and {Lin}, Su-Jie and {Wang}, Bing-Bing and {Yin}, Peng-Fei},
	title = "{Excesses of cosmic ray spectra from a single nearby source}",
	journal = {prd},
	year = 2017,
	month = jul,
	volume = {96},
	number = {2},
	eid = {023006},
	pages = {023006},
	doi = {10.1103/PhysRevD.96.023006},
	archivePrefix = {arXiv},
	eprint = {1611.09118},
	primaryClass = {astro-ph.HE},
	adsurl = {https://ui.adsabs.harvard.edu/abs/2017PhRvD..96b3006L}
}

@ARTICLE{2007Ap&SS.308..225F,
	author = {{Faherty}, Jacqueline and {Walter}, Frederick M. and {Anderson}, Jay},
	title = "{The trigonometric parallax of the neutron star Geminga}",
	journal = {\apss},
	year = 2007,
	month = apr,
	volume = {308},
	number = {1-4},
	pages = {225-230},
	doi = {10.1007/s10509-007-9368-0},
	adsurl = {https://ui.adsabs.harvard.edu/abs/2007Ap&SS.308..225F}
}

@Article{2012ApJ...752L..13T,
	Title                    = {{Origin of the Cosmic-Ray Spectral Hardening}},
	Author                   = {{Tomassetti}, N.},
	Journal                  = {apj},
	Year                     = {2012},
	
	Month                    = jun,
	Pages                    = {L13},
	Volume                   = {752},
	Adsurl                   = {http://adsabs.harvard.edu/abs/2012ApJ...752L..13T},
	Archiveprefix            = {arXiv},
	Doi                      = {10.1088/2041-8205/752/1/L13},
	Eid                      = {L13},
	Eprint                   = {1204.4492},
	Primaryclass             = {astro-ph.HE}
}

@ARTICLE{2016ApJ...819...54G,
	author = {{Guo}, Y.-Q. and {Tian}, Z. and {Jin}, C.},
	title = "{Spatial-dependent Propagation of Cosmic Rays Results in the Spectrum of Proton, Ratios of P/P, and B/C, and Anisotropy of Nuclei}",
	journal = {\apj},
	year = 2016,
	month = mar,
	volume = 819,
	eid = {54},
	pages = {54},
	doi = {10.3847/0004-637X/819/1/54},
	adsurl = {http://adsabs.harvard.edu/abs/2016ApJ...819...54G}
}

@ARTICLE{2018PhRvD..97f3008G,
	author = {{Guo}, Yi-Qing and {Yuan}, Qiang},
	title = "{Understanding the spectral hardenings and radial distribution of Galactic cosmic rays and Fermi diffuse {\ensuremath{\gamma}} rays with spatially-dependent propagation}",
	journal = {\prd},
	year = 2018,
	month = mar,
	volume = {97},
	number = {6},
	eid = {063008},
	pages = {063008},
	doi = {10.1103/PhysRevD.97.063008},
	archivePrefix = {arXiv},
	eprint = {1801.05904},
	primaryClass = {astro-ph.HE},
	adsurl = {https://ui.adsabs.harvard.edu/abs/2018PhRvD..97f3008G}
}

@ARTICLE{2020ApJ...892....6L,
	author = {{Liu}, Wei and {Lin}, Su-jie and {Hu}, Hong-bo and {Guo}, Yi-qing and {Li}, Ai-feng},
	title = "{Two Numerical Methods for the 3D Anisotropic Propagation of Galactic Cosmic Rays}",
	journal = {\apj},
	year = 2020,
	month = mar,
	volume = {892},
	number = {1},
	eid = {6},
	pages = {6},
	doi = {10.3847/1538-4357/ab765a},
	archivePrefix = {arXiv},
	eprint = {1909.02908},
	primaryClass = {astro-ph.HE},
	adsurl = {https://ui.adsabs.harvard.edu/abs/2020ApJ...892....6L}
}

@ARTICLE{2019JCAP...10..010L,
	author = {{Liu}, Wei and {Guo}, Yi-Qing and {Yuan}, Qiang},
	title = "{Indication of nearby source signatures of cosmic rays from energy spectra and anisotropies}",
	journal = {jcap},
	year = 2019,
	month = oct,
	volume = {2019},
	number = {10},
	eid = {010},
	pages = {010},
	doi = {10.1088/1475-7516/2019/10/010},
	archivePrefix = {arXiv},
	eprint = {1812.09673},
	primaryClass = {astro-ph.HE},
	adsurl = {https://ui.adsabs.harvard.edu/abs/2019JCAP...10..010L}
}

@ARTICLE{2019JCAP...12..007Q,
	author = {{Qiao}, Bing-Qiang and {Liu}, Wei and {Guo}, Yi-Qing and {Yuan}, Qiang},
	title = "{Anisotropies of different mass compositions of cosmic rays}",
	journal = {jcap},
	year = 2019,
	month = dec,
	volume = {2019},
	number = {12},
	eid = {007},
	pages = {007},
	doi = {10.1088/1475-7516/2019/12/007},
	archivePrefix = {arXiv},
	eprint = {1905.12505},
	primaryClass = {astro-ph.HE},
	adsurl = {https://ui.adsabs.harvard.edu/abs/2019JCAP...12..007Q}
}

@ARTICLE{2017Sci...358..911A,
	author = {{Abeysekara}, A.~U. and {Albert}, A. and {Alfaro}, R. and {Alvarez}, C. and
	{{\'A}lvarez}, J.~D. and {Arceo}, R. and
	{Arteaga-Vel{\'a}zquez}, J.~C. and {Avila Rojas}, D. and
	{Ayala Solares}, H.~A. and {Barber}, A.~S. and {Bautista-Elivar}, N. and
	{Becerril}, A. and {Belmont-Moreno}, E. and {BenZvi}, S.~Y. and
	{Berley}, D. and {Bernal}, A. and {Braun}, J. and {Brisbois}, C. and
	{Caballero-Mora}, K.~S. and {Capistr{\'a}n}, T. and
	{Carrami{\~n}ana}, A. and {Casanova}, S. and {Castillo}, M. and
	{Cotti}, U. and {Cotzomi}, J. and {Couti{\~n}o de Le{\'o}n}, S. and
	{De Le{\'o}n}, C. and {De la Fuente}, E. and {Dingus}, B.~L. and
	{DuVernois}, M.~A. and {D{\'\i}az-V{\'e}lez}, J.~C. and
	{Ellsworth}, R.~W. and {Engel}, K. and {Enr{\'\i}quez-Rivera}, O. and
	{Fiorino}, D.~W. and {Fraija}, N. and
	{Garc{\'\i}a-Gonz{\'a}lez}, J.~A. and {Garfias}, F. and {Gerhardt}, M. and
	{Gonz{\'a}lez Mu{\~n}oz}, A. and {Gonz{\'a}lez}, M.~M. and
	{Goodman}, J.~A. and {Hampel-Arias}, Z. and {Harding}, J.~P. and
	{Hern{\'a}ndez}, S. and {Hern{\'a}ndez-Almada}, A. and {Hinton}, J. and
	{Hona}, B. and {Hui}, C.~M. and {H{\"u}ntemeyer}, P. and {Iriarte}, A. and
	{Jardin-Blicq}, A. and {Joshi}, V. and {Kaufmann}, S. and {Kieda}, D. and
	{Lara}, A. and {Lauer}, R.~J. and {Lee}, W.~H. and {Lennarz}, D. and
	{Vargas}, H. Le{\'o}n and {Linnemann}, J.~T. and {Longinotti}, A.~L. and
	{Luis Raya}, G. and {Luna-Garc{\'\i}a}, R. and {L{\'o}pez-Coto}, R. and
	{Malone}, K. and {Marinelli}, S.~S. and {Martinez}, O. and
	{Martinez-Castellanos}, I. and {Mart{\'\i}nez-Castro}, J. and
	{Mart{\'\i}nez-Huerta}, H. and {Matthews}, J.~A. and {Mirand
	a-Romagnoli}, P. and {Moreno}, E. and {Mostaf{\'a}}, M. and
	{Nellen}, L. and {Newbold}, M. and {Nisa}, M.~U. and
	{Noriega-Papaqui}, R. and {Pelayo}, R. and {Pretz}, J. and
	{P{\'e}rez-P{\'e}rez}, E.~G. and {Ren}, Z. and {Rho}, C.~D. and
	{Rivi{\`e}re}, C. and {Rosa-Gonz{\'a}lez}, D. and {Rosenberg}, M. and
	{Ruiz-Velasco}, E. and {Salazar}, H. and {Salesa Greus}, F. and {Sand
	oval}, A. and {Schneider}, M. and {Schoorlemmer}, H. and {Sinnis}, G. and
	{Smith}, A.~J. and {Springer}, R.~W. and {Surajbali}, P. and
	{Taboada}, I. and {Tibolla}, O. and {Tollefson}, K. and {Torres}, I. and
	{Ukwatta}, T.~N. and {Vianello}, G. and {Weisgarber}, T. and
	{Westerhoff}, S. and {Wisher}, I.~G. and {Wood}, J. and {Yapici}, T. and
	{Yodh}, G. and {Younk}, P.~W. and {Zepeda}, A. and {Zhou}, H. and
	{Guo}, F. and {Hahn}, J. and {Li}, H. and {Zhang}, H.},
	title = "{Extended gamma-ray sources around pulsars constrain the origin of the positron flux at Earth}",
	journal = {Science},
	year = 2017,
	month = nov,
	volume = {358},
	number = {6365},
	pages = {911-914},
	doi = {10.1126/science.aan4880},
	archivePrefix = {arXiv},
	eprint = {1711.06223},
	primaryClass = {astro-ph.HE},
	adsurl = {https://ui.adsabs.harvard.edu/abs/2017Sci...358..911A}
}

@ARTICLE{2013ApJ...776...30F,
	author = {{Funsten}, H.~O. and {DeMajistre}, R. and {Frisch}, P.~C. and
	{Heerikhuisen}, J. and {Higdon}, D.~M. and {Janzen}, P. and
	{Larsen}, B.~A. and {Livadiotis}, G. and {McComas}, D.~J. and
	{M{\"o}bius}, E. and {Reese}, C.~S. and {Reisenfeld}, D.~B. and
	{Schwadron}, N.~A. and {Zirnstein}, E.~J.},
	title = "{Circularity of the Interstellar Boundary Explorer Ribbon of Enhanced Energetic Neutral Atom (ENA) Flux}",
	journal = {\apj},
	year = "2013",
	month = "Oct",
	volume = {776},
	number = {1},
	eid = {30},
	pages = {30},
	doi = {10.1088/0004-637X/776/1/30},
	adsurl = {https://ui.adsabs.harvard.edu/abs/2013ApJ...776...30F}
}

@ARTICLE{2016PhRvL.117o1103A,
	author = {{Ahlers}, M.},
	title = "{Deciphering the Dipole Anisotropy of Galactic Cosmic Rays}",
	journal = {Physical Review Letters},
	archivePrefix = "arXiv",
	eprint = {1605.06446},
	primaryClass = "astro-ph.HE",
	year = 2016,
	month = oct,
	volume = 117,
	number = 15,
	eid = {151103},
	pages = {151103},
	doi = {10.1103/PhysRevLett.117.151103},
	adsurl = {http://adsabs.harvard.edu/abs/2016PhRvL.117o1103A}
}

@ARTICLE{2017PrPNP..94..184A,
	author = {{Ahlers}, Markus and {Mertsch}, Philipp},
	title = "{Origin of small-scale anisotropies in Galactic cosmic rays}",
	journal = {Progress in Particle and Nuclear Physics},
	year = 2017,
	month = may,
	volume = {94},
	pages = {184-216},
	doi = {10.1016/j.ppnp.2017.01.004},
	archivePrefix = {arXiv},
	eprint = {1612.01873},
	primaryClass = {astro-ph.HE},
	adsurl = {https://ui.adsabs.harvard.edu/abs/2017PrPNP..94..184A}
}

@ARTICLE{2015ApJ...809...90B,
	author = {{Bartoli}, B. and {Bernardini}, P. and {Bi}, X.~J. and {Cao}, Z. and 
	{Catalanotti}, S. and {Chen}, S.~Z. and {Chen}, T.~L. and {Cui}, S.~W. and 
	{Dai}, B.~Z. and {D'Amone}, A. and {Danzengluobu} and {De Mitri}, I. and 
	{D'Ettorre Piazzoli}, B. and {Di Girolamo}, T. and {Di Sciascio}, G. and 
	{Feng}, C.~F. and {Feng}, Z. and {Feng}, Z. and {Gao}, W. and 
	{Gou}, Q.~B. and {Guo}, Y.~Q. and {He}, H.~H. and {Hu}, H. and 
	{Hu}, H. and {Iacovacci}, M. and {Iuppa}, R. and {Jia}, H.~Y. and 
	{Labaciren} and {Li}, H.~J. and {Liu}, C. and {Liu}, J. and 
	{Liu}, M.~Y. and {Lu}, H. and {Ma}, L.~L. and {Ma}, X.~H. and 
	{Mancarella}, G. and {Mari}, S.~M. and {Marsella}, G. and {Mastroianni}, S. and 
	{Montini}, P. and {Ning}, C.~C. and {Perrone}, L. and {Pistilli}, P. and 
	{Salvini}, P. and {Santonico}, R. and {Shen}, P.~R. and {Sheng}, X.~D. and 
	{Shi}, F. and {Surdo}, A. and {Tan}, Y.~H. and {Vallania}, P. and 
	{Vernetto}, S. and {Vigorito}, C. and {Wang}, H. and {Wu}, C.~Y. and 
	{Wu}, H.~R. and {Xue}, L. and {Yang}, Q.~Y. and {Yang}, X.~C. and 
	{Yao}, Z.~G. and {Yuan}, A.~F. and {Zha}, M. and {Zhang}, H.~M. and 
	{Zhang}, L. and {Zhang}, X.~Y. and {Zhang}, Y. and {Zhao}, J. and 
	{Zhaxiciren} and {Zhaxisangzhu} and {Zhou}, X.~X. and {Zhu}, F.~R. and 
	{Zhu}, Q.~Q. and {ARGO-YBJ Collaboration}},
	title = "{ARGO-YBJ Observation of the Large-scale Cosmic Ray Anisotropy During the Solar Minimum between Cycles 23 and 24}",
	journal = {apj},
	year = 2015,
	month = aug,
	volume = 809,
	eid = {90},
	pages = {90},
	doi = {10.1088/0004-637X/809/1/90},
	adsurl = {http://adsabs.harvard.edu/abs/2015ApJ...809...90B}
}

@ARTICLE{2017JCAP...10..019C,
	author = {{Cerri}, S.~S. and {Gaggero}, D. and {Vittino}, A. and {Evoli}, C. and 
	{Grasso}, D.},
	title = "{A signature of anisotropic cosmic-ray transport in the gamma-ray sky}",
	journal = {jcap},
	archivePrefix = "arXiv",
	eprint = {1707.07694},
	primaryClass = "astro-ph.HE",
	year = 2017,
	month = oct,
	volume = 10,
	eid = {019},
	pages = {019},
	doi = {10.1088/1475-7516/2017/10/019},
	adsurl = {https://ui.adsabs.harvard.edu/abs/2017JCAP...10..019C}
}

@ARTICLE{1996A&AS..120C.437C,
	author = {{Case}, G. and {Bhattacharya}, D.},
	title = "{Revisiting the galactic supernova remnant distribution.}",
	journal = {\aaps},
	year = 1996,
	month = dec,
	volume = {120},
	pages = {437-440},
	adsurl = {https://ui.adsabs.harvard.edu/abs/1996A&AS..120C.437C}
}

@ARTICLE{2007ARNPS..57..285S,
	author = {{Strong}, Andrew W. and {Moskalenko}, Igor V. and {Ptuskin}, Vladimir S.},
	title = "{Cosmic-Ray Propagation and Interactions in the Galaxy}",
	journal = {Annual Review of Nuclear and Particle Science},
	year = 2007,
	month = nov,
	volume = {57},
	number = {1},
	pages = {285-327},
	doi = {10.1146/annurev.nucl.57.090506.123011},
	archivePrefix = {arXiv},
	eprint = {astro-ph/0701517},
	primaryClass = {astro-ph},
	adsurl = {https://ui.adsabs.harvard.edu/abs/2007ARNPS..57..285S}
}

@ARTICLE{2020NatAs...4.1001Z,
       author = {{Zhang}, Heshou and {Chepurnov}, Alexey and {Yan}, Huirong and {Makwana}, Kirit and {Santos-Lima}, Reinaldo and {Appleby}, Sarah},
        title = "{Identification of plasma modes in Galactic turbulence with synchrotron polarization}",
      journal = {Nature Astronomy},
         year = 2020,
        month = may,
       volume = {4},
        pages = {1001-1008},
          doi = {10.1038/s41550-020-1093-4},
archivePrefix = {arXiv},
       eprint = {1808.01913},
 primaryClass = {physics.plasm-ph},
       adsurl = {https://ui.adsabs.harvard.edu/abs/2020NatAs...4.1001Z}
}

@ARTICLE{2020PhRvX..10c1021M,
       author = {{Makwana}, K.~D. and {Yan}, Huirong},
        title = "{Properties of Magnetohydrodynamic Modes in Compressively Driven Plasma Turbulence}",
      journal = {Physical Review X},
         year = 2020,
        month = jul,
       volume = {10},
       number = {3},
          eid = {031021},
        pages = {031021},
          doi = {10.1103/PhysRevX.10.031021},
archivePrefix = {arXiv},
       eprint = {1907.01853},
 primaryClass = {physics.plasm-ph},
       adsurl = {https://ui.adsabs.harvard.edu/abs/2020PhRvX..10c1021M}
}

@article{Liu:2025Pr,
  author = "Liu, Wei  and  Li, Dan  and  Sun, Dong-xu  and  Hu, Hong-bo  and  Yuan, Qiang  and  Guo, Yi-Qing",
  title = "{Measurements of cosmic-ray anisotropy using LHAASO-WCDA}",
  doi = "10.22323/1.501.0320",
  journal = "PoS",
  year = 2025,
  volume = "ICRC2025",
  pages = "320"
}

@ARTICLE{1999ApJ...520..204G,
 	author = {{Giacalone}, J. and {Jokipii}, J.~R.},
 	title = "{The Transport of Cosmic Rays across a Turbulent Magnetic Field}",
 	journal = {\apj},
 	year = 1999,
 	month = jul,
 	volume = 520,
 	pages = {204-214},
 	doi = {10.1086/307452},
 	adsurl = {https://ui.adsabs.harvard.edu/abs/1999ApJ...520..204G}
 }

@ARTICLE{1966ApJ...146..480J,
	author = {{Jokipii}, J.~R.},
	title = "{Cosmic-Ray Propagation. I. Charged Particles in a Random Magnetic Field}",
	journal = {\apj},
	year = 1966,
	month = nov,
	volume = {146},
	pages = {480},
	doi = {10.1086/148912},
	adsurl = {https://ui.adsabs.harvard.edu/abs/1966ApJ...146..480J}
}

@ARTICLE{2022PhRvL.129y1103A,
       author = {{Adriani}, O. and {Akaike}, Y. and {Asano}, K. and {Asaoka}, Y. and {Berti}, E. and {Bigongiari}, G. and {Binns}, W.~R. and {Bongi}, M. and {Brogi}, P. and {Bruno}, A. and {Buckley}, J.~H. and {Cannady}, N. and {Castellini}, G. and {Checchia}, C. and {Cherry}, M.~L. and {Collazuol}, G. and {de Nolfo}, G.~A. and {Ebisawa}, K. and {Ficklin}, A.~W. and {Fuke}, H. and {Gonzi}, S. and {Guzik}, T.~G. and {Hams}, T. and {Hibino}, K. and {Ichimura}, M. and {Ioka}, K. and {Ishizaki}, W. and {Israel}, M.~H. and {Kasahara}, K. and {Kataoka}, J. and {Kataoka}, R. and {Katayose}, Y. and {Kato}, C. and {Kawanaka}, N. and {Kawakubo}, Y. and {Kobayashi}, K. and {Kohri}, K. and {Krawczynski}, H.~S. and {Krizmanic}, J.~F. and {Maestro}, P. and {Marrocchesi}, P.~S. and {Messineo}, A.~M. and {Mitchell}, J.~W. and {Miyake}, S. and {Moiseev}, A.~A. and {Mori}, M. and {Mori}, N. and {Motz}, H.~M. and {Munakata}, K. and {Nakahira}, S. and {Nishimura}, J. and {Okuno}, S. and {Ormes}, J.~F. and {Ozawa}, S. and {Pacini}, L. and {Papini}, P. and {Rauch}, B.~F. and {Ricciarini}, S.~B. and {Sakai}, K. and {Sakamoto}, T. and {Sasaki}, M. and {Shimizu}, Y. and {Shiomi}, A. and {Spillantini}, P. and {Stolzi}, F. and {Sugita}, S. and {Sulaj}, A. and {Takita}, M. and {Tamura}, T. and {Terasawa}, T. and {Torii}, S. and {Tsunesada}, Y. and {Uchihori}, Y. and {Vannuccini}, E. and {Wefel}, J.~P. and {Yamaoka}, K. and {Yanagita}, S. and {Yoshida}, A. and {Yoshida}, K. and {Zober}, W.~V. and {Calet Collaboration}},
        title = "{Cosmic-Ray Boron Flux Measured from 8.4 GeV /n to 3.8 TeV /n with the Calorimetric Electron Telescope on the International Space Station}",
      journal = {\prl},
         year = 2022,
        month = dec,
       volume = {129},
       number = {25},
          eid = {251103},
        pages = {251103},
          doi = {10.1103/PhysRevLett.129.251103},
archivePrefix = {arXiv},
       eprint = {2212.07873},
 primaryClass = {astro-ph.HE},
       adsurl = {https://ui.adsabs.harvard.edu/abs/2022PhRvL.129y1103A}
}

@ARTICLE{2017ApJ...839....5Y,
	author = {{Yoon}, Y.~S. and {Anderson}, T. and {Barrau}, A. and {Conklin}, N.~B. and {Coutu}, S. and {Derome}, L. and {Han}, J.~H. and {Jeon}, J.~A. and {Kim}, K.~C. and {Kim}, M.~H. and {Lee}, H.~Y. and {Lee}, J. and {Lee}, M.~H. and {Lee}, S.~E. and {Link}, J.~T. and {Menchaca-Rocha}, A. and {Mitchell}, J.~W. and {Mognet}, S.~I. and {Nutter}, S. and {Park}, I.~H. and {Picot-Clemente}, N. and {Putze}, A. and {Seo}, E.~S. and {Smith}, J. and {Wu}, J.},
	title = "{Proton and Helium Spectra from the CREAM-III Flight}",
	journal = {\apj},
	year = 2017,
	month = apr,
	volume = {839},
	number = {1},
	eid = {5},
	pages = {5},
	doi = {10.3847/1538-4357/aa68e4},
	archivePrefix = {arXiv},
	eprint = {1704.02512},
	primaryClass = {astro-ph.HE},
	adsurl = {https://ui.adsabs.harvard.edu/abs/2017ApJ...839....5Y}
}

@ARTICLE{1996ApJ...463..224P,
       author = {{Plucinsky}, Paul P. and {Snowden}, Steven L. and {Aschenbach}, Bernd and {Egger}, Roland and {Edgar}, Richard J. and {McCammon}, Dan},
        title = "{ROSAT Survey Observations of the Monogem Ring}",
      journal = {\apj},
         year = 1996,
        month = may,
       volume = {463},
        pages = {224},
          doi = {10.1086/177236},
       adsurl = {https://ui.adsabs.harvard.edu/abs/1996ApJ...463..224P}
}

@ARTICLE{2022ApJ...926...41Z,
       author = {{Zhao}, Bing and {Liu}, Wei and {Yuan}, Qiang and {Hu}, Hong-Bo and {Bi}, Xiao-Jun and {Wu}, Han-Rong and {Zhou}, Xun-Xiu and {Guo}, Yi-Qing},
        title = "{Geminga SNR: Possible Candidate of the Local Cosmic-Ray Factory}",
      journal = {\apj},
         year = 2022,
        month = feb,
       volume = {926},
       number = {1},
          eid = {41},
        pages = {41},
          doi = {10.3847/1538-4357/ac4416},
archivePrefix = {arXiv},
       eprint = {2104.07321},
 primaryClass = {astro-ph.HE},
       adsurl = {https://ui.adsabs.harvard.edu/abs/2022ApJ...926...41Z}
}

\end{document}